\documentclass[useAMS,usenatbib]{mn2e}
\usepackage{graphicx}
\usepackage{color}

\newcommand{\rad}{$\rm R_{\rm e}$}

\newcommand{\ls}{$\log \sigma$}

\newcommand{\simlt}{\lower.5ex\hbox{$\; \buildrel < \over \sim \;$}}
\newcommand{\simgt}{\lower.5ex\hbox{$\; \buildrel > \over \sim \;$}}

\newcommand{\mgfep}{$\rm [MgFe]'$}
\newcommand{\mgfe}{$\rm [MgFe]$}

\newcommand{\hbo}{$\rm H\beta_o$}
\newcommand{\hb}{$\rm H\beta$}

\newcommand{\hgf}{$\rm H\gamma_F$}

\newcommand{\kms}{\,km\,s$^{-1}$}
\newcommand{\afe}{[$\alpha$/Fe]}
\newcommand{\afep}{$[Z_{\rm Mg}/Z_{\rm Fe}]$}
\newcommand{\zh}{$[Z/H]$}
\newcommand{\mh}{M$_{\rm h}$}
\newcommand{\lmh}{$\log($M$_{\rm h}/$M$_\odot)$}
\newcommand{\lmha}{$\log($M$_{\rm h,1}/$M$_\odot)$}
\newcommand{\lmhb}{$\log($M$_{\rm h,2}/$M$_\odot)$}
\newcommand{\av}{$\rm A_V$}
\newcommand{\betaafe}{$\beta_{\alpha \! / \! {\rm Fe}}$}
\newcommand{\betazh}{$\beta_{Z \! / \! H}$}
\newcommand{\mstar}{$\rm M_\star$}
\newcommand{\dfib}{$\rm D_{fiber}$}
\title[The environmental fossil record of ETGs]
{SPIDER X -- Environmental effects in central and satellite early-type galaxies 
through the stellar fossil record}
\author[La Barbera et al.]
{F. La Barbera$^{1}$\thanks{E-mail: labarber@na.astro.it (FLB)}, 
A. Pasquali$^{2}$, I. Ferreras$^{3}$, A. Gallazzi$^{4,5}$, R.R. de
Carvalho$^{6}$, \newauthor I.G. de la Rosa$^{7,8}$
\\ 
$^{1}$ INAF -- Osservatorio Astronomico di Capodimonte, Salita
Moiariello 16, 80020 Napoli, Italy\\
$^{2}$ Universit\"at Heidelberg, Zentrum fur Astronomie,
Astronomisches Rechen-Institut, M\"onchhofstr. 12-14, D-69120
Heidelberg, Germany\\
$^{3}$ Mullard Space Science Laboratory, University College London,
Holmbury St Mary, Dorking, Surrey RH5 6NT, UK\\
$^{4}$ INAF-Osservatorio Astrofisico di Arcetri, Largo Enrico Fermi 5,
50125 Firenze, Italy\\
$^{5}$ Dark Cosmology Center, University of Copenhagen, Niels Bohr
Institute, Juliane Maries Vej 30, 2100 Copenhagen, Denmark\\
$^{6}$ Instituto Nacional de Pesquisas Espaciais/MCT,
  S. J. dos Campos, Brazil\\
$^{7}$ Instituto de Astrof\'\i sica de Canarias (IAC), E-38200
  La Laguna, Tenerife, Spain\\
$^{8}$ Departamento de Astrof\'\i sica, Universidad de La
  Laguna, E-38205, Tenerife, Spain
}
\begin{document}

\date{Revised version, \today}

\pagerange{\pageref{firstpage}--\pageref{lastpage}} \pubyear{2014}

\maketitle

\label{firstpage}

\begin{abstract}
A detailed analysis of how environment affects the star formation
history of early-type galaxies (ETGs) is undertaken via high signal to
noise ratio stacked spectra obtained from a sample of $20,977$ ETGs
(morphologically selected) from the SDSS-based SPIDER survey.  Two
major parameters are considered for the study: the central velocity
dispersion ($\sigma$), which relates to local drivers of star
formation, and the mass of the host halo, which relates to
environment-related effects.  In addition, we separate the sample
between centrals (the most massive galaxy in a halo) and satellites.
We derive trends of age, metallicity, and \afe\ enhancement, with
$\sigma$.  We confirm that the major driver of stellar population
properties in ETGs is velocity dispersion, with a second-order effect
associated to the central/satellite nature of the galaxy. No
environmental dependence is detected for satellite ETGs, except at low
$\sigma$ -- where satellites in groups or in the outskirts of
clusters tend to be younger than those in the central regions of
clusters. In contrast, the trends for centrals show a significant
dependence on halo mass.  Central ETGs in groups (i.e.  with a halo
mass $>10^{12.5}M_\odot$) have younger ages, lower \afe , and higher
internal reddening, than ``isolated'' systems (i.e. centrals residing
in low-mass, $<10^{12.5}M_\odot$, halos).  Our findings imply that
central ETGs in groups formed their stellar component over longer time
scales than ``isolated'' centrals, mainly because of gas-rich
interactions with their companion galaxies.
\end{abstract}


%%%%%%%%%%%%%%%%%%%%%%%%%%%%%%%%%%%%%%%%%%%%%%%%%%%%%%%%%%%%%%%%%
\section{Introduction}
\label{sec:intro}

Despite  many  observational and  theoretical  efforts  over the  past
thirty years,  the population of early-type galaxies  (ETGs) remains a
challenge of  modern cosmology.  Galaxy formation and  evolution is an
extremely   variegated  discipline,  involving   complex  ``internal''
processes,  such as  the transformation  of gas  into  stars, feedback
processes from, e.g., active  galactic nuclei and supernov\ae, as well
as  ``external'' processes,  such as  interactions among  galaxies and
between galaxies  and their environment  (e.g.  ram-pressure stripping
induced  by the  hot intra-cluster  medium). Despite  this complexity,
most  of the  observed  properties of  ETGs  appear very  homogeneous,
showing  tight  correlations with  a  single  key observable,  central
velocity    dispersion    \citep[see,   e.g.][]{Bernardi:03}.     Such
correlations involve  structural parameters (namely,  the scale radius
and  mean  luminosity  density  within  that radius),  also  known  as
Fundamental Plane relation~\citep{Djorgovski:87, Dressler:87}, as well
as  the  underlying  stellar   populations  (e.g.   age  and  chemical
composition).

In the last twenty years,  our knowledge of the stellar populations of
ETGs  has advanced, thanks  to improvements  in the  spectroscopic and
photometric    observations    available    \citep[e.g.,][]{Trager:00,
  Thomas:05, Gallazzi:06,  Bernardi:05, Smith:07, Pasquali:10, Zhu:10,
  Thomas:10,  Harrison:11}, as  well  as advances  in the  theoretical
modelling  of  unresolved  stellar  populations  \citep[e.g.,][]{wo94,
  BC:03,Maraston:05,Vazdekis:2012}.   It is  now widely  accepted that
both  metallicity  and  the  abundance of  the  $\alpha$-elements  (in
particular  the ratio  [Mg/Fe]) increase  with galaxy  mass,  the main
driver of  these correlations  being the central  velocity dispersion,
$\sigma$    (i.e.     the    central    potential     well)    of    a
galaxy~\citep{Bernardi:03}, or  its dynamical mass~\citep{Gallazzi:06,
  Chang:06}.   The   stellar  metallicity-$\sigma$  relation   can  be
understood because most of the massive systems retain more efficiently
the  heavy elements  from exploding  supernov\ae\ within  their deeper
potential well~\citep{Larson:74,ArY:87}.   Since $\alpha$ elements are
released during  the explosion of  short-lived massive stars  (type II
supernov\ae), in contrast to iron,  that is mainly produced by type Ia
supernov\ae, the  ratio \afe\ is believed  to serve as a  proxy of the
star-formation time scale of  a galaxy.  Therefore, the \afe -$\sigma$
relation can be interpreted as a trend towards shorter, more efficient
star    formation    in     the    most    massive    ETGs~\citep[see,
  e.g.,][]{Thomas:99,Thomas:05}. Regarding stellar ages, early studies
found      no       significant      correlation      with      galaxy
mass~(e.g.~\citealt{Trager:00, Kunt:01}), while more recent works have
found the age to increase with central velocity dispersion, especially
at low  velocity dispersion.  The  trends of age and  metallicity with
mass  are  not limited  to  the central  regions  of  ETGs, probed  by
spectroscopic   studies,    but   also   apply    to   their   overall
(i.e. aperture-integrated)  stellar populations, as implied  by the FP
relation~\citep{SpiderII}.   In  addition  to  age,  metallicity,  and
abundance ratios,  the (unresolved)  stellar populations of  ETGs have
also been  recently found  to feature a  varying stellar  Initial Mass
Function (IMF), through the  analysis of gravity-sensitive features in
their  spectra. A  significant  correlation is  also  found in  place,
between IMF  slope and velocity  dispersion, with massive  ETGs having
enhanced  fractions   of  dwarf-to-giant  stars  with   respect  to  a
``standard'',  Kroupa-like   IMF~\citep{Cenarro2001,  vDC:10,  CvD12b,
  Ferreras:13, LaBarbera:13, Spiniello:14}.

Although galaxy mass (or more precisely, velocity dispersion) is found
to be the primary driver of the stellar population properties of ETGs,
these properties  are also affected by the  environment where galaxies
reside.   Several  studies  have   analysed  the  dependence  of  age,
metallicity, and \afe , on ``local'' environment, characterised either
by the density of neighbouring galaxies, the distance to the centre
of  the  host  group/cluster,  or  by splitting  galaxies  into  those
residing   in    galaxy   groups   and   those    in   the   ``field''
(e.g.~\citealt{Bernardi:03,     Thomas:05,    BERN:06,    Annibali:07,
  Clemens:09,  Thomas:10,  Pasquali:10,  Zhu:10,  Cooper:10}).   These
works have all found that galaxies in low-density regions have younger
luminosity-weighted  ages than  their cluster  counterparts,  with age
differences  of  $\sim  1$--$2$~Gyr (e.g.~\citealt{GLC:92,  Trager:00,
  Kunt:02, TF:02, Thomas:05,  BERN:06, Clemens:09}).  Such differences
also  exist  when   considering  ETGs  in  high-density,  low-velocity
dispersion     systems,     such     as    the     Hickson     Compact
Groups~\citep{deLaRosa:07}. In contrast,  discordant results have been
reported in  the literature  regarding trends between  environment and
metallicity.   \citet{BERN:06} found no  detectable difference  in the
metallicity  of  field  and   cluster  ETGs,  whereas  other  studies,
e.g.~\citet{Thomas:05,  deLaRosa:07,  Clemens:09}  found evidence  for
galaxies in dense environments to be less metal rich than those in the
field. One  possible cause of  these discrepancies is  the significant
dependence  of   the  internal   metallicity  gradients  of   ETGs  on
environment (see~\citealt{LaBarbera:11}  and references therein), with
different  studies  targeting  different  apertures  for  the  stellar
population analysis.   The environmental dependence of  \afe\ has been
addressed less  frequently in the literature than  age and metallicity
trends, mostly because of the lack of stellar population models, based
on  {\sl empirical}  stellar  libraries  that cover  a  range in  age,
metallicity, and \afe . \citet{BERN:06}  found ETGs at high density to
be more $\alpha$-enhanced  than those at low density,  indicative of a
shorter formation  time scale of their  stars. \citet{Thomas:05} found
that the  fraction of ``rejuvenated''  objects, having young  ages and
lower  \afe, increases  from high-  to low-density  regions.  However,
these trends with environment are biased by galaxy stellar mass, as it
is well known that more massive galaxies inhabit denser regions of the
Universe \citep{dressler:80}.  Such a coupling  between galaxy stellar
mass and environment  has to be carefully taken  into account in order
to   single  out   the  effects   of  environment   alone  \citep[see,
  e.g.,][]{Pasquali:09,Cooper:10}.

Although local density is an intuitive way to describe the environment
where a galaxy resides, it is not necessarily the most effective way
to characterise it.  In the current picture (e.g.~\citealt{Croton:06,
  delucia:06}), two main paths of galaxy formation and evolution can
be envisioned.  Galaxies can be either ``centrals'' in a surrounding
dark-matter halo, or ``satellites'', being accreted onto the halo
during its hierarchical build-up. Centrals can keep accreting hot gas
from the halos they are embedded in, whereas satellites are believed
to be deprived of their gas reservoir by their host halo.  Therefore,
splitting galaxies into centrals and satellites is a more natural way
to compare observations and theory.  Another interesting approach to
characterise the environment is that of using the shape of the
velocity distribution of galaxies in a group (see,
e.g.,~\citealt{Ribeiro:2013}).  Despite the fact that several studies
have analysed the dependence of stellar populations of ETGs on local
environment, there has been no detailed stellar population study of
the two populations of satellite and central ETGs individually.  This
is indeed the main focus of the present work.  By analysing stacked
spectra obtained from a large sample of morphologically selected ETGs
from the SDSS-based SPIDER survey~\citep{SpiderI}, we explore how the
stellar populations of central and satellite ETGs -- namely the age,
metallicity, \afe, and internal extinction -- depend on the ``global''
environment (i.e. the mass of the parent halo) where they reside.  We
compare galaxy properties at fixed velocity dispersion, in order to
single out the effect of galaxy mass and environment.  The present
analysis complements ~\citet[hereafter PGF10]{Pasquali:10}, differing
from it in several regards: (i) we analyse here only the population of
morphologically selected ETGs; (ii) the properties of galaxies are
compared at fixed velocity dispersion, rather than fixed galaxy
stellar mass; (iii) we study here also the environmental dependence of
\afe\ and internal extinction, in addition to age and metallicity (as
in PGF10).  Throughout the present work, we adopt a cosmology with
$\Omega_{\rm m}=0.3$, $\Omega_{\Lambda}=0.7$, and 
H$_0 = 100h$\,\kms\,Mpc$^{-1}$, with $h=0.7$.

%%%%%%%%%%%%%%%%%%%%%%%%%%%%%%%%%%%%%%%%%%%%%%%%%%%%%%%%%%%%%%%%%
\section{Data}
\label{sec:data}

%%%%%%%%%%%%%%%%%%%%%%%%%%%%%%%%%%%%%%%%%%%%%%%%%%%%%%%%%%%%%%%%%
\subsection{Sample}
\label{sec:sample}

The SPIDER\footnote{Spheroids  Panchromatic Investigation in Different
  Environmental  Regions~\citep{SpiderI}}  sample  consists of  39,993
nearby  (0.05$<$z$<$0.095) ETGs selected  from Data  Release 6  of the
Sloan  Digital Sky  Survey  (SDSS-DR6;~\citealt{SDSS:DR6}).  Following
~\citet{BER03},  ETGs  are  defined  as bulge-dominated  systems  with
passive  spectra. The  former is  constrained via  the  SDSS attribute
$fracDev_r \!  > \!  0.8$,  where $fracDev_r$ measures the fraction of
galaxy light better fitted by a de~Vaucouleurs profile, rather than an
exponential law. For the latter,  we use the SDSS attribute $eClass \!
< \!  0$, where $eClass$ indicates the spectral type of a galaxy based
on  a  principal   component  analysis  decomposition.   Galaxies  are
selected to have absolute  $r$-band Petrosian magnitudes brighter than
$r_{Petr}=-20$, corresponding  to the observed  separation between the
two    families     of    {\it    bright}     and    {\it    ordinary}
ellipticals~\citep{capaccioli1992, graham&guzman2003}.  As detailed in
\citet[][hereafter Paper I]{SpiderI}, for  a subsample of $5,080$ ETGs
there is also NIR  ($YJHK$) photometry available from UKIDSS-DR3.  All
galaxy images, in the  $grizYJHK$ wavebands, were homogeneously fitted
with   two-dimensional   PSF-convolved   S\'ersic  models,   providing
structural parameters, namely the  effective radius, \rad, the S\'ersic
(shape-parameter),  $n$, and  the total  luminosity, from  $g$ through
$K$.   All  SPIDER ETGs  have  spectra  as  well as  central  velocity
dispersions, $\sigma$, available from  SDSS. The spectra, ranging from
$3800$     to     $9200$\,\AA,     have    been     retrieved     from
SDSS-DR7~\citep{SDSS:DR7},  de-redshifted to  a common  rest-frame and
corrected for foreground Galactic extinction (see Paper I).

For the present study, we rely on a subsample of SPIDER ETGs\footnote{
 We notice that although using the entire SDSS-DR7 would provide
better statistics than the SPIDER SDSS subsample, this choice
benefits from the detailed characterization of this sample, as presented in our
previous papers (e.g. morphological selection, estimates of stellar
masses and structural parameters, characterization of the environment
using different group catalogues). Moreover, enlarging the redshift
range would bring further issues into the analysis (e.g. aperture
effects, incompleteness at low $\sigma$), that we want to
minimize.},
selected as follows. Following \citet[][hereafter Paper
  VIII]{LaBarbera:13}, we select objects with: (i) $\sigma>100$\,\kms
($N_{ETGs}=38,447$); (ii) low internal extinction 
($E(B \!  - \!V)<0.1$~mag\footnote{ 
As detailed in Paper VIII, the $E(B-V)$ has been estimated with
the spectral fitting code STARLIGHT for each individual spectrum,
following a similar procedure to that described in
Sec.~\ref{sec:age_zh} for the stacked spectra { in this paper}.};
$N_{ETGs}=33,095$); and (iii) spectra whose $S/N$
ratio computed per \AA\/ in the H$\beta$ region is higher than 14, 27
and 21 at $\sigma$ = 100, 200 and 300\,\kms, respectively (thus
excluding those objects within the lowest quartile of the $S/N$
distribution in a given $\sigma$ bin; $N_{ETGs}=24,781$).  Since ETGs
in the SPIDER sample are defined through the SDSS $fracDev_r$
parameter, which is a proxy of bulge fraction, the sample is affected
by some amount of contamination (up to $17\%$, see Paper I) from
late-type galaxies (LTGs) with a prominent bulge.  Due to the
morphology--density relation~\citep{dressler:80}, the contaminant
fraction is expected to increase towards lower density regions,
potentially introducing a systematic trend in the study of stellar
populations as a function of environment.  

To avoid this, we define a sample of bona-fide, morphologically
selected, ETGs. Firstly, we take advantage of the publicly available
catalogue of the Galaxy Zoo project, that provides the morphological
classification of nearly 900,000 SDSS galaxies~\citep{lintott:2011}.
Except for a few cases (72 out of $39,993$ galaxies), inspected
visually by ourselves, SPIDER ETGs have a Galaxy Zoo classification.
We use the {\it fraction of votes} for ellipticals, {\it p$ _{ell}$},
spirals (both clock-wise and anti-clock-wise), {\it p$_{sp}$}, and
edge-on spirals, {\it p$_{edge}$}. Face-on LTGs have {\it p$_{sp}$}
$>$ {\it p$_{ell}$}, while edge-on disks have {\it p$_{edge}$} $>$
{\it p$_{ell}$}. Under these criteria, our main SPIDER sample contains
$1,806$ face-on LTGs, $3,545$ edge-on LTGs (4.5 and 8.9\,\% of the
total, respectively), and $712$ objects (1.8\,\%) tagged as
unclassified/mergers.  Since some early-type spirals (Sa/SBa) are not
present in the Galaxy Zoo classification, we apply a further selection
criterion based on the quality of the two-dimensional S\'ersic fits to
the surface brightness distribution. For each fit, the $\chi^2$ is
measured as the rms of the residuals from the fit.  High values of
$\chi^2$ usually correlate with the presence of faint morphological
features, such as disks or spiral arms (see, e.g., figure~6 of Paper
I). We use the fitting results for the g-band images, which are
expected to be sensitive to the presence of young stellar
components. As expected, the $\chi^2$ correlates with the Galaxy Zoo
classification. Edge-on and face-on LTGs have a median
$\langle\chi^2\rangle$ of $1.31$ and $1.41$, respectively, while the
remaining sample has a median $\langle\chi^2\rangle=1.02$ (consistent
with figure~5 of Paper I), with a standard deviation
$\sigma_{\chi^2}=0.17$.  We define as residual contaminants those
objects not flagged as LTGs from the Galaxy Zoo classification, but
residing in the high-end tail of the $\chi^2$ distribution, i.e.
$\chi^2 > \langle\chi^2\rangle + 3 \sigma_{\chi^2}$. This selection
excluded an additional $852$ objects (2.1\,\%) of the main sample. Out
of $24,781$ galaxies selected with the same criteria as in Paper VIII
(see points i--iii above), $3,126$ are removed after the morphological
selection described above, resulting in a sample of $21,655$
bona-fide ETGs.  Finally, $20,977$ (out of $21,655$) ETGs have
environment defined, and are binned into five subsamples of ETGs
residing in different environments, as detailed below.

{ The SPIDER sample is approximately complete down to $r_{Petr} =
  -20$, which is one of the main criteria imposed to select ETGs (see
  above, and Paper I for details).  Because of the scatter in the
  $\sigma$--luminosity relation, this $r$-band selection may induce some
  incompleteness at low velocity dispersion. To quantify it, we have
  estimated the fraction of ETGs excluded by the $r_{Petr}$ cut,
  defined as $f_{>r}$, as a function of $\sigma$. We find $f_{>r}\sim
  10\%$ at $\sigma \sim 140$\,\kms , increasing up to $\sim 25\%$ at
  $110$\,\kms. We notice that $f_{>r}$ is an upper estimate of
  incompleteness, as galaxies fainter than $-20$ are {\it ordinary}
  rather than {\it bright} ETGs (see below), i.e. do not necessarily
  belong to the population of galaxies we want to target.  Finally, we
  remark that although incompleteness might affect the low-$\sigma$
  trends of stellar population properties, it does not affect the
  relative comparison of these properties with environment at fixed
  $\sigma$, which is the main goal of this paper.}

%%%%%%%%%%%%%%%%%%%%%%%%%%%%%%%%%%%%%%%%%%%%%%%%%%%%%%%%%%%%%%%%%
\subsection{Environment}
\label{sec:environment}

We  characterise the  environment where  ETGs reside  by means  of the
updated    catalogue     of    galaxy    groups    of~\citet[hereafter
  Y07]{yang:2007}. The difference between  the updated version and the
one defined in  Y07 is the area used,  i.e.  SDSS-DR7 \citep{SDSS:DR7}
rather  than DR4  \citep{SDSS:DR4}.  The  catalogue is  constructed by
applying the halo-based group finder algorithm of~\citet{yang:2005} to
the    New    York    University    Value-Added    Galaxy    Catalogue
(NYU-VAGC,~\citealt{blanton:2005}) extracted  from SDSS-DR7.  From the
NYU-VAGC  Main  Galaxy Sample,  Y07  extracted  all  galaxies with  an
$r$-band  apparent  magnitude brighter  than  $r$  =  18 mag,  in  the
redshift  interval  0.01  $\leq  z  \leq$ 0.20  and  with  a  redshift
completeness $C_z >$  0.7. Three samples were built  with the selected
galaxies:  sample  I, which  only  uses  the  $593,736$ galaxies  with
measured redshifts  from the  SDSS; sample II  based on  the $593,736$
galaxies with measured redshifts from SDSS, plus an additional $3,115$
galaxies with  SDSS photometry  and redshifts from  different sources;
sample III which lists $36,602$ additional galaxies lacking a redshift
due  to fibre  collisions,  but  being assigned  the  redshift of  the
nearest  neighbour.  For  our analysis,  we rely  on sample  II, where
galaxies are  split into {\it  centrals} (defined as the  most massive
group  members  on  the  basis   of  galaxy  stellar  mass)  and  {\it
  satellites}  (all group members  that are  not centrals).   The dark
matter halo mass, M$_{\rm h}$, provided in this sample for each galaxy
group, is based on the ranking  of the group total stellar mass, where
galaxy  stellar  masses  are  derived from  the  relationship  between
stellar  mass-to-light ratio  and colour  of Bell  et  al.~(2003). The
method of Y07 can only assign  halo masses to groups more massive than
M$_{\rm  h}\sim$10$^{12}~h^{-1}$M$_{\odot}$,  and  with  one  or  more
members brighter than M$_{\rm r}  - 5\log h = -19.5$\,mag (K-corrected
to $z$  = 0.1). For M$_{\rm h}\simgt$10$^{12}~h^{-1}$M$_{\odot}$, the
Y07 group finder successfully selects more than 90$\%$ of all ``true''
halos (see Y07 for details),  almost independent of their richness and
with  only a  very weak  dependence on  halo mass.   For  less massive
groups, Yang, Mo \& van den Bosch (2008) used the relationship between
the stellar mass of central galaxies and the halo mass of their groups
to estimate the  halo mass of single central  galaxies down to M$_{\rm
  h} \sim$ 10$^{11}~h^{-1}$M$_{\odot}$.   Y07 assessed the uncertainty
on  \mh\ by  measuring  halo masses  for  a mock  group catalogue  and
comparing  them  to true  dark-matter  halo  masses.   The scatter  on
\mh\  has  been found  to  range from  $\sim  0.35$\,dex  at \mh  $\sim
13.5$--$14$  to   $\sim  0.2$\,dex   at  high-  (\mh$\sim   14.6$)  and
low-(\mh$\sim  12$) halo masses  (see their  fig.~7).  We  discuss the
possible   effect  of  the   \mh\  uncertainty   on  our   results  in
Sec.~\ref{sec:trends}.

\begin{figure}
\begin{center}
\leavevmode
\includegraphics[width=8cm]{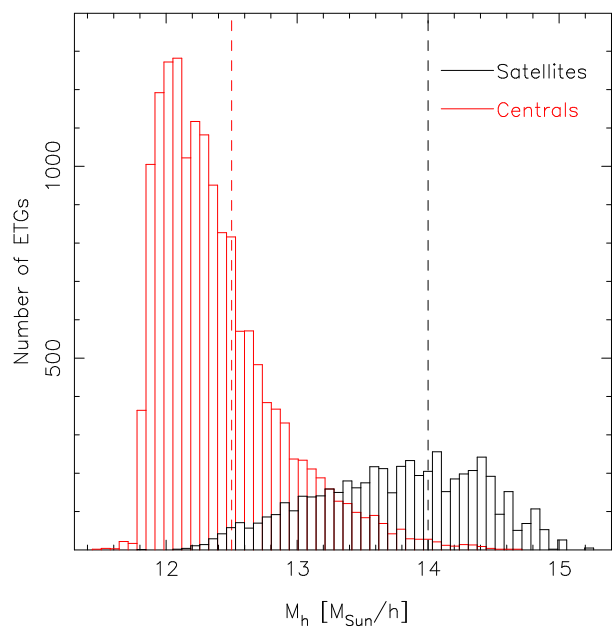}
\end{center}
\caption{(Logarithmic) halo mass distribution of our sample of
  $20,977$ bona-fide ETGs. The histograms of galaxies classified as
  centrals and satellites in the updated Y07 group catalogue (see
  Sec.~\ref{sec:environment}) are shown in red and black,
  respectively.  The vertical dashed lines mark the threshold values
  of the halo mass that split centrals (red) and satellites (black)
  (see the text for details).  }
%\contcaption{.}
\label{fig:hmass_hist}
\end{figure}

Out of  $21,655$ bona-fide  ETGs selected for  the present  study (see
above), $20,977$ ETGs are in sample II of the Y07 group catalogue, and
are   thus   classified  as   centrals   ($N=15,572$)  or   satellites
($N=5,425$). Fig.~\ref{fig:hmass_hist} shows  the mass distribution of
the parent halos  where ETGs reside. Satellites span  a wider range of
halo  mass, from  \lmh$\sim  13$, i.e.  the  mass scale  of groups  of
galaxies,  to \lmh$\sim 15$,  i.e.  massive  galaxy clusters.   On the
contrary,  most centrals  ($\sim 89  \,  \%$) have  \lmh\ below  $13$,
reflecting the higher  number density of low- to  high-mass halos. For
\lmh $<12.5$, centrals are  indeed ``isolated'' galaxies, as suggested
by the  fact that  only a minor  fraction, $\sim 2\,\%$,  of satellite
ETGs  reside  in halos  with  masses  below  this value.  In  general,
considering the entire sample II of  Y07 we found $<9\%$ of all groups
with \mh$<12.5$  to have (both early- and  late-type) satellites, with
only $1\%$  of them having more  than one satellite.   Hence, we split
central ETGs into two subsamples, having low halo mass ($<12.5$), that
can be  regarded as ``isolated''  galaxies (hereafter sample  C1), and
high  halo mass  ($\ge 12.5$),  i.e. groups'  central  ETGs (hereafter
sample  C2),  respectively  (see  red--dashed  vertical  line  in  the
Figure).  These two subsamples include $68 \, \%$ ($N=10,534$) and $32
\, \%$ ($N=5,038$) of all central ETGs ($N=15,572$), respectively.  In
the satellite ETG population,  we define three subsamples, namely, two
subsamples  comprising low-  (\lmh  $<14$; $N=2,945$,  sample S1)  and
high-  (\lmh $  \ge 14$;  $N=1,582$,  sample S2)  halo-mass, at  small
cluster-centric  projected  radii  ($R\le  0.5 \,  R_{200}$);  and  an
additional third  subsample consisting of satellite ETGs  in the outer
group regions  ($R> 0.5  \, R_{200}$), regardless  of group  halo mass
($N=898$,  sample S3).   Notice that  S3 is  too small  for  a further
subdivision  according to  halo mass.   To simplify  our  notation, we
refer to satellites  at low and high halo mass  as group- and cluster-
satellites, respectively.   In summary,  we define five  subsamples of
ETGs  residing  in  different  environments,  namely  four  subsamples
(either centrals or satellites) defined with respect to halo mass, and
an additional subsample of satellites  that lie in the external regions
of the host  halo. Notice that we adopt a  different threshold in halo
mass to split  centrals and satellites, as our goal  is not to compare
satellites with centrals,  but to analyse the effect  of the halo mass
on each population separately. The general details of these subsamples
are summarised in Tab.~\ref{tab:samples}.

\begin{table}
\small
\centering
\begin{minipage}{90mm}
 \caption{Samples  of ETGs  residing in  different  environments. Each
   sample  is  identified by  a  label  (col.~1).   The definition  of
   environment is summarised in  col.~2, while col.~3 gives the sample
   size.  }
  \begin{tabular}{c|c|r}
 \hline
  sample & environment & $N$ \\
    (1) & (2) & (3) \\
 \hline
 $\rm C1$ & centrals,   \lmh$<  12.5$ & $10,534$ \\
 $\rm C2$ & centrals,   \lmh$\ge 12.5$ & $5,038$ \\
 $\rm S1$ & satellites, $R/R_{200} < 0.5$, \lmh$< 14$ & $2,945$ \\
 $\rm S2$ & satellites, $R/R_{200} < 0.5$, \lmh$\ge 14$ & $1,582$ \\
 $\rm S3$ & satellites, $R/R_{200} \ge 0.5$ & $898$ \\
 \hline
\end{tabular}
\label{tab:samples}
\end{minipage}
\end{table}

\begin{table*}
\centering
\small
\begin{minipage}{170mm}
 \caption{Properties of  stacked spectra for samples  of ETGs residing
   in different  environments (see labels  in Tab.~\ref{tab:samples}),
   i.e. central  ETGs with low-  and high- host halo  mass (cols.~1--3
   and 4--6, respectively), and  satellite ETGs with low- (cols.~7--9)
   and high- (cols.~10--12) \mh , and at large projected group-centric
   distances  (cols.~13--15).  For  each sample,  the Table  lists the
   velocity  dispersion  bins  with  their $\sigma$  range  (in  \kms,
   cols.~1, 4,  7, 10, 13), the number  of ETGs, $\rm N$,  in each bin
   (col.~2,  5, 8,  11,  14), and  the  $S/N$ ratio  per  \AA\ of  the
   corresponding stacked spectrum (cols.~3, 6, 9, 12, 15).}
%  \begin{tabular}{l|c|r||l|c|r|l|c|r|l|c|r|l|c|r}
  \begin{tabular}{ccc|ccc|ccc|ccc|ccc}
   \hline
   \hline
   \multicolumn{3}{c}{C1} & \multicolumn{3}{c}{C2} & \multicolumn{3}{c}{S1} & 
   \multicolumn{3}{c}{S2} & \multicolumn{3}{c}{S3} \\
   \hline
$\sigma$ & $\rm N$ & $S/N$ & $\sigma$ & $\rm N$ & $S/N$ & $\sigma$ & $\rm N$ & $S/N$ & $\sigma$ & $\rm N$ & $S/N$ & $\sigma$ & $\rm N$ & $S/N$ \\
(1)&(2)&(3)&(4)&(5)&(6)&(7)&(8)&(9)&(10)&(11)&(12)&(13)&(14)&(15)\\
\hline
	$100$--$110$ & $ 630$ & $ 371$ &$100$--$130$ & $ 101$ & $ 180$ &	$100$--$110$ & $ 173$ & $ 194$ &	$100$--$110$ & $  53$ & $ 117$ &	$100$--$110$ & $  52$ & $ 107$ \\
	$110$--$120$ & $1093$ & $ 493$ &$130$--$140$ & $  94$ & $ 174$ &	$110$--$120$ & $ 264$ & $ 238$ &	$110$--$120$ & $ 110$ & $ 158$ &	$110$--$120$ & $  73$ & $ 131$ \\
	$120$--$130$ & $1387$ & $ 590$ &$140$--$150$ & $ 189$ & $ 233$ &	$120$--$130$ & $ 321$ & $ 281$ &	$120$--$130$ & $ 139$ & $ 190$ &	$120$--$130$ & $ 108$ & $ 169$ \\
	$130$--$140$ & $1422$ & $ 609$ &$150$--$160$ & $ 274$ & $ 289$ &	$130$--$140$ & $ 359$ & $ 312$ &	$130$--$140$ & $ 145$ & $ 186$ &	$130$--$140$ & $ 115$ & $ 185$ \\
	$140$--$150$ & $1430$ & $ 624$ &$160$--$170$ & $ 402$ & $ 354$ &	$140$--$150$ & $ 329$ & $ 306$ &	$140$--$150$ & $ 146$ & $ 214$ &	$140$--$150$ & $ 110$ & $ 184$ \\
	$150$--$160$ & $1256$ & $ 617$ &$170$--$180$ & $ 473$ & $ 403$ &	$150$--$160$ & $ 300$ & $ 308$ &	$150$--$160$ & $ 139$ & $ 196$ &	$150$--$160$ & $  91$ & $ 161$ \\
	$160$--$170$ & $1150$ & $ 617$ &$180$--$190$ & $ 532$ & $ 428$ &	$160$--$170$ & $ 285$ & $ 312$ &	$160$--$170$ & $ 156$ & $ 244$ &	$160$--$170$ & $  92$ & $ 197$ \\
	$170$--$180$ & $ 793$ & $ 516$ &$190$--$200$ & $ 526$ & $ 443$ &	$170$--$180$ & $ 216$ & $ 288$ &	$170$--$180$ & $ 142$ & $ 230$ &	$170$--$180$ & $  53$ & $ 161$ \\
	$180$--$190$ & $ 566$ & $ 447$ &$200$--$210$ & $ 554$ & $ 471$ &	$180$--$190$ & $ 195$ & $ 279$ &	$180$--$190$ & $ 115$ & $ 214$ &	$180$--$190$ & $  53$ & $ 141$ \\
	$190$--$200$ & $ 356$ & $ 372$ &$210$--$220$ & $ 446$ & $ 440$ &	$190$--$200$ & $ 158$ & $ 252$ &	$190$--$200$ & $ 109$ & $ 206$ &	$190$--$210$ & $  70$ & $ 205$ \\
	$200$--$210$ & $ 194$ & $ 277$ &$220$--$230$ & $ 393$ & $ 438$ &	$200$--$210$ & $ 112$ & $ 229$ &	$200$--$210$ & $  78$ & $ 200$ &	$210$--$230$ & $  42$ & $ 148$ \\
	$210$--$220$ & $ 123$ & $ 210$ &$230$--$240$ & $ 335$ & $ 423$ &	$210$--$220$ & $  86$ & $ 220$ &	$210$--$220$ & $  65$ & $ 169$ &	         $ $ &   $ $  & $ $    \\
	$220$--$230$ & $  59$ & $ 160$ &$240$--$250$ & $ 234$ & $ 341$ &	$220$--$230$ & $  58$ & $ 159$ &	$220$--$230$ & $  55$ & $ 165$ &	         $ $ &   $ $  & $ $    \\
	$230$--$250$ & $  58$ & $ 177$ &$250$--$260$ & $ 174$ & $ 299$ &	$230$--$250$ & $  60$ & $ 203$ &	$230$--$250$ & $  63$ & $ 222$ &	         $ $ &   $ $  & $ $    \\
	         $ $ &   $ $  & $ $    &$260$--$270$ & $ 118$ & $ 251$ &	         $ $ &   $ $  & $ $    &	$250$--$280$ & $  51$ & $ 211$ &	         $ $ &   $ $  & $ $    \\
	         $ $ &   $ $  & $ $    &$270$--$280$ & $  72$ & $ 223$ &	         $ $ &   $ $  & $ $    &	         $ $ &   $ $  & $ $    &	         $ $ &   $ $  & $ $    \\
	         $ $ &   $ $  & $ $    &$280$--$290$ & $  57$ & $ 180$ &	         $ $ &   $ $  & $ $    &	         $ $ &   $ $  & $ $    &	         $ $ &   $ $  & $ $    \\
	         $ $ &   $ $  & $ $    &$290$--$310$ & $  52$ & $ 206$ &	         $ $ &   $ $  & $ $    &	         $ $ &   $ $  & $ $    &	         $ $ &   $ $  & $ $    \\
  \hline
\end{tabular}
\label{tab:stacks}
\end{minipage}
\end{table*}

%%%%%%%%%%%%%%%%%%%%%%%%%%%%%%%%%%%%%%%%%%%%%%%%%%%%%%%%%%%%%%%%%
\subsection{Stacked spectra}
\label{sec:stacks}
Following  the same  procedure  as in  Paper  VIII, the  spectroscopic
analysis of the  sample is based on stacked  spectra, providing a high
$S/N$  ($>100$\,\AA$^{-1}$).  For each  environment-related subsample,
the spectra are stacked in  bins of velocity dispersion, starting from
$\sigma=100$\,\kms  ,  with a  minimum  bin  size, $\delta_\sigma$,  of
10\,\kms . This size is chosen to match the error on the measurement of
the  velocity dispersion over  the range  100\,\kms $\leq  \sigma \leq$
300\,\kms.  When necessary, the bin size, $\delta_\sigma$, is increased
adaptively, in steps of $10$\,\kms , up to a maximum value of $30$\,\kms
, in order to have a minimum of $40$ spectra per bin. Bins that do not
fulfil  these constraints  (i.e.  $N\ge40$  for  $\delta_{\sigma} \le
30$\,\kms  ) are  not included  in  the analysis.   Adopting a  minimum
number of  $40$ spectra in  each bin ensures  that the $S/N$  ratio is
large for  all stacked spectra.  A  maximum bin size  of $30$\,\kms\ is
adopted because  the aim  of this  work is to  quantify the  effect of
``global''  (large-scale) environment  with respect  to  the ``local''
(galaxy scale) driver of the  star formation histories of ETGs. A good
proxy    for    the   latter    is    central   velocity    dispersion
\citep{Bernardi:05}.  The  spectra in each $\sigma$  bin are convolved
to match the  upper $\sigma$ limit of the  bin. Subsequently, for each
sample, and each $\sigma$ bin,  we median stack the available spectra,
considering only pixels with  no SDSS flag raised\footnote{i.e. no bad
  pixels,      flat     field      issues,      etc;     see      {\tt
    http://www.sdss.org/dr6/dm/flatFiles/spSpec.html}               for
  details}. Noise in stacked spectra is computed as the uncertainty of
median  flux values, accounting  for the  actual distribution  of flux
values at  each wavelength.  This procedure  results in a  total of 72
stacked spectra, corresponding to  central and satellite ETGs residing
in different  environments, and over a velocity  dispersion range from
$\sim 100$ to $\sim 250$\,\kms\ (with the upper value depending on the
sample).   Relevant  properties  of   all  stacks  are  summarised  in
Tab.~\ref{tab:stacks}, where we report the range of each $\sigma$ bin,
the number of ETGs per bin,  and the median $S/N$ ratio of the stacks,
computed in the central passband  of the age-sensitive \hb\ line (from
$4840$ to $4880$ \AA ).  The stacked spectra feature a remarkably high
$S/N$, larger than  $\sim 100$\AA$^{-1}$ for all bins,  with a maximum
of  $\sim  300$\AA$^{-1}$ for  central  ETGs  at  low halo  mass  (see
Tab.~\ref{tab:stacks}).

%%%%%%%%%%%%%%%%%%%%%%%%%%%%%%%%%%%%%%%%%%%%%%%%%%%%%%%%%%%%%%%%%
\section{Stellar population properties}
\label{sec:sp_pars}
This paper  focuses on estimating stellar  population properties, i.e.
$Age$,  total  metallicity,   \zh  ,  and  $\alpha$-to-iron  abundance
variations, \afe, using high  $S/N$ stacked spectra to robustly derive
the  contribution  of   galaxy-scale  (via  velocity  dispersion)  and
halo-scale (via halo mass) physics  to the star formation histories of
ETGs. In the  following, we describe the procedure  followed to obtain
$Age$        and        \zh\       (Sec.~\ref{sec:age_zh}),        and
\afe\ (Sec.~\ref{sec:alpha}).

%%%%%%%%%%%%%%%%%%%%%%%%%%%%%%%%%%%%%%%%%%%%%%%%%%%%%%%%%%%%%%%%%
\subsection{Age and metallicity}
\label{sec:age_zh}

\begin{figure*}
\begin{center}
\leavevmode
\includegraphics[width=14cm]{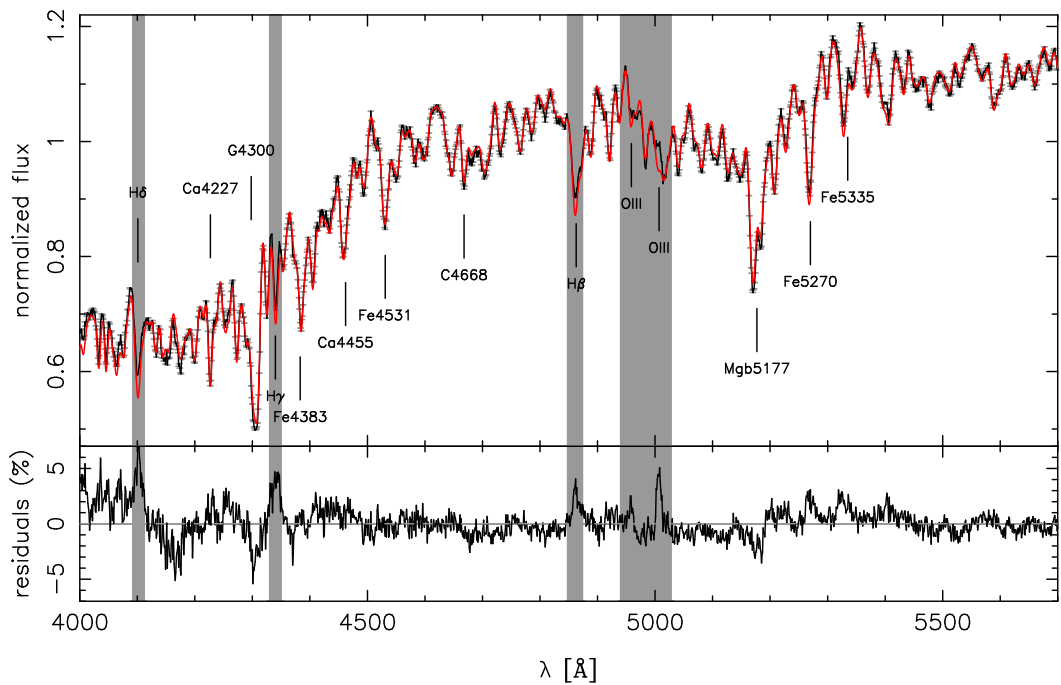}
\end{center}
\caption{ Example of a STARLIGHT fit for one of our stacked spectra,
  drawn from sample C2, with $\sigma \sim 200$\,\kms . The upper panel
  shows the stacked spectrum (black curve), the STARLIGHT best-fit
  (red curve), one-sigma uncertainties of the flux values (grey error bars).
  The regions excluded from the fit because of possible contamination from
  nebular emission are shaded in grey. Some prominent absorption
  features are marked with vertical ticks and labelled. The lower panel
  plots the relative residuals with respect to the fit.
%  \textcolor{red}{Mebbe we should also show the shaded grey regions in 
%  the lower panel to illustrate where the residuals are high?
%  }
}
%\contcaption{.}
\label{fig:starlight}
\end{figure*}

We  fit each  stacked spectrum  with  the spectral  fitting code  {\tt
  STARLIGHT}~\citep{CID05}. For a  given input observed spectrum, {\tt
  STARLIGHT} determines the best-fitting linear combination of a basis
set  of models.   The  basis comprises  $N_{\rm  SSP}$ simple  stellar
populations (hereafter SSP). Each SSP  is characterised by a fixed age
and chemical  composition. The best  fitting solution is  described by
the relative light-weighted contribution, $\{x_j\}\ (j=1,\cdots,N_{\rm
  SSP})$,  that corresponds  to each  of the  SSPs in  the  basis.  In
addition,  {\tt  STARLIGHT}  provides  an  estimate  of  the  internal
reddening, \av , of the input spectrum, by including an extinction law
in the fitting process.  We adopt  here a set of $108$ input SSPs from
the  MILES  galaxy   spectral  library~\citep{Vazdekis10},  with  $27$
log-spaced  ages  ranging  from   $0.5$  to  $\sim  17.78$\,Gyr,  four
metallicities  $[Z/H]=\{-0.71,  -0.40,  0.00,+0.22\}$,  and  a  Kroupa
Universal Initial Mass Function  (IMF).  MILES SSPs cover the spectral
range   $3525$--$7500$\,\AA,  with  a   spectral  resolution   of  2.3
\AA\ \citep{Jesus:11} .   Hence, they are well suited  to analyse SDSS
spectra,  whose  spectral  resolution  is $\sim$  2.3\,\AA\  FWHM,  at
$\lambda \sim  5350$\,\AA\ (see, e.g., Paper VIII).   MILES models are
based on stellar spectra in the solar neighbourhood (the MILES stellar
library\footnote{\tt www.iac.es/proyecto/miles}, see~\citealt{SB06a}),
which are  solar-scaled only  at solar and  super-solar metallicities,
having \afe  $>0$ at  low metallicities ($[Z/H]<-0.4$).   In practice,
this deviation from solar-scale is not relevant for the present study,
as even in our lowest  $\sigma$ stacks, the metallicities are slightly
sub-solar (\zh $> -0.15$).  Notice  also that although we include SSPs
in our basis  with ages older than the current estimate  of the age of
Universe  \citep[13.75\,Gyr,][]{WMAP9},  the  derived $Age$  estimates
from {\tt  STARLIGHT} do  not exceed $11$\,Gyr  for any of  our stacks
(see  below), i.e.  consistent  with the  age of  the Universe  at the
median redshift ($z \sim 0.07$) of our sample of ETGs ($\sim 12.5$~Gyr
with the  adopted cosmology). More importantly, we  emphasise that the
zero-point  $Age$  calibration of  stellar  population models  remains
uncertain at present, suggesting  that relative (rather than absolute)
age differences  should be considered in  stellar population synthesis
studies. Regarding  internal extinction, a~\citet{Cardelli} extinction
law   is  adopted,   suitable   for  systems   with   low  levels   of
star-formation,  like ETGs.   Spectral fitting  is performed  over the
spectral range from  $4000$ to $5700$\,\AA . We  exclude regions bluer
than $4000$\,\AA\ as they are more sensitive to (i) small fractions of
young  stars in  a stellar  population, and  (ii)  non-solar abundance
ratios  of chemical elements  in the  stellar atmospheres,  that might
bias  the estimate  of age  and metallicity  when relying  on (nearly)
solar-scaled (MILES) stellar population  models.  We also exclude from
the  fitting window  those regions  potentially affected  by (nebular)
emission,   i.e.    $H{\delta}$   ($4092$--$4112$\,\AA),   $H{\gamma}$
($4330$--$4350$\,\AA), $H{\beta}$  ($4848$--$4874$\,\AA), and $[OIII]$
($4940$--$5028$\,\AA).  Wavelengths longer  than $5700$\,\AA\ are also
excluded from  the fitting as  they include several features  (e.g. Na
absorption at $5900$  and $8200$\,\AA, as well as  TiO bands) that are
especially  sensitive  to  the  stellar  IMF (Paper  VIII).   While  a
systematic  variation of  the IMF  with velocity  dispersion  has been
detected  in ETGs  (e.g.~\citealt{CvD12a,  Ferreras:13, Spiniello:14};
Paper VIII), we leave the issue of a possible dependence of the IMF on
environment to a  future paper. To allow for  a better comparison with
previous  works,  we  derive  the correlation  of  stellar  population
properties with velocity dispersion  assuming a constant, Kroupa, IMF.
{ Fig.~\ref{fig:starlight} shows an example of the best-fitting
  spectrum produced by STARLIGHT, for one of our stacks (corresponding
  to sample C2, with $\sigma \sim 200$\,\kms ), with the regions
  excluded from the fit in grey. Residuals are at the
  level of a few percent throughout the whole spectral range fitted.}

For each  stack, we run {\tt  STARLIGHT} by first  smoothing the basis
SSP models  to match the  wavelength-dependent resolution of  the SDSS
spectrograph  (see Paper  VIII for  details) as  well as  the velocity
dispersion of  the stack  under study.  Hence,  for a  given property,
$Y$,   (either   age   or   metallicity),   we   estimate   its   {\it
  luminosity-weighted} value, $Y_L$, as
\begin{equation}
Y_L = \frac{\sum Y \times x_j}{\sum x_j},
\label{eq:lum_weight_SP}
\end{equation}
where  the sum  extends  over all  the  basis SSPs.  \zh\ is  obtained
directly from Eq.~\ref{eq:lum_weight_SP},  whereas for $Age$, we first
estimate the  logarithmic luminosity-weighted value,  and then convert
it to  linear units. The uncertainties  on $Age$ and  \zh\ are derived
via  a bootstrap  procedure, where  {\tt STARLIGHT}  is run  for $200$
realisations of  the input spectrum, randomly modifying  each time the
flux values according to their  uncertainties. The errors on $Age$ and
\zh\ are given as the  widths of the distributions of these parameters
among the different realisations. Due to the high $S/N$ of the stacks,
the typical  random error on $Age$  is $\sim 2$\%, while  for \zh\ the
typical  uncertainty amounts  to $\sim  0.006$\,dex  (computing median
values over all choices of $\sigma$/environment).

We notice that a potentially  important issue for the present study is
that  we estimate $Age$  and \zh\  with (nearly)  solar-scaled stellar
population  models.  For  the  $\alpha$-enhanced  stellar  populations
expected in  massive ETGs, the  use of solar-scaled models  might bias
the inferred  estimate of age  and metallicity. More  importantly, any
dependence of  \afe\ on environment might also  introduce a systematic
variation  of age and  metallicity --  when derived  with solar-scaled
models.  Since at the moment there is no set of SSP models, {\it based
  on  empirical  stellar  spectra},  that  take  \afe\  properly  into
account, one way to overcome this problem would be to estimate age and
metallicity  from single spectral  features (like  \hb\ and  the total
metallicity indicator \mgfep ), that are believed to be independent of
\afe.  The  main drawback of this  approach is that one  does not take
full  advantage  of the  information  encoded  in  a galaxy  spectrum,
hampering a detailed study of  the SFH. Based on these considerations,
in the  present work we  adopt the following approach.   Our reference
estimates of age and metallicity  are those obtained with widely used,
well   referenced,    solar-scaled   MILES   models.     However,   in
App.~\ref{app:afe_ssps}, we also adopt  a preliminary version of MILES
SSPs with  varying \afe\ (\citealt{Cervantes07};  based on theoretical
spectral  libraries),  to re-derive  age  and  metallicity, and  their
trends with $\sigma$ and  environment.  Our main results, as discussed
throughout the paper,  are robust, and do not depend  on the choice of
adopted  models.  Furthermore,  we also  compare  line-strengths among
different stacked spectra, finding  very consistent results with those
from    {\tt   STARLIGHT}    (Sec.~\ref{sec:ews}).    As    shown   in
Appendix~\ref{app:compAG}, our results  are also consistent with those
obtained    with    a   different    approach    (similar   to    that
of~\citealt{Gallazzi:06}), where  median stellar population properties
are  derived  from  a  Bayesian  analysis of  absorption  features  on
individual galaxies (Gallazzi et al., in prep).

%%%%%%%%%%%%%%%%%%%%%%%%%%%%%%%%%%%%%%%%%%%%%%%%%%%%%%%%%%%%%%%%%
\subsection{Alpha enhancement}
\label{sec:alpha}

We estimate a solar-scale proxy for \afe , following the same approach
as in Paper VIII. Given the luminosity-weighted age obtained with {\tt
  STARLIGHT}, we obtain two independent metallicity estimates for each
spectrum, $Z_{Mg}$  and $Z_{Fe}$, using  the spectral indices  Mgb and
${\rm    Fe}3\equiv({\rm   Fe}4383+{\rm    Fe}5270+{\rm   Fe}5335)/3$,
respectively.  These  metallicities   are  derived  by  comparing  the
equivalent widths  (EWs) of  either Mgb or  Fe3 to predictions  of the
MILES SSP models  with fixed age, as illustrated  in Figure~5 of Paper
VIII. While at solar scale both $[Z/H]_{Mg}$ and $[Z/H]_{Fe}$ would be
the  same, for  an  $\alpha$-enhanced population  the $[Z/H]_{Mg}$  is
larger than $[Z/H]_{Fe}$. Since  the $[Z/H]_{Mg}$ is often larger than
the maximum  metallicity of MILES SSPs  ($[Z/H]=+0.22$), the procedure
also involves an  extrapolation of the model EWs to  \zh $>$ 0.22 (see
Paper  VIII).  The  solar proxy  of \afe\  is defined  as \afep$\equiv
[Z/H]_{Mg}-[Z/H]_{Fe}$.  As shown  in  Paper VIII,  the  \afep\ has  a
remarkably  tight correlation  with  estimates of  \afe\ from  stellar
population   models   taking   abundance   ratios   explicitly   into
account~\citep{TMJ11}, and is also robust against the approach used to
estimate the  age (i.e. a comparison between  {\tt STARLIGHT} spectral
fitting and  an analysis of  Balmer line strengths).  For  all stacked
spectra,  we translate  \afep\ into  \afe\ using  the  above mentioned
correlation  from  Paper VIII,  i.e.  \afe  $\sim  0.55 \cdot$\afep  .
Notice that we  adopt the same procedure to estimate  $Age$, \zh , and
\afe\  for all  stacks,  allowing for  a  meaningful comparison  among
different environments and $\sigma$ bins. Following the same procedure
as  for $Age$  and \zh\  ,  the uncertainties  on \afe\  (\afep )  are
obtained with  a bootstrap approach,  where line strengths as  well as
age values are shifted according to their uncertainties, and the width
of the \afe\ distribution is computed among $200$ iterations.

%%%%%%%%%%%%%%%%%%%%%%%%%%%%%%%%%%%%%%%%%%%%%%%%%%%%%%%%%%%%%%%%%
\section{Results}
\label{sec:results}

We compare  the trends of age,  metallicity, and \afe\  among the five
samples    of     ETGs    residing    in     different    environments
(Sec.~\ref{sec:environment}) as a function of velocity dispersion.  In
Sec.~\ref{sec:trends},  we  show the  results  obtained with  spectral
fitting ({\tt STARLIGHT}; Sec.~\ref{sec:age_zh}) and our proxy of \afe
(Sec.~\ref{sec:alpha}), while  Sec.~\ref{sec:ews} tests the robustness
of the results, comparing the  trends of line strengths with $\sigma$,
for specific spectral features.

%%%%%%%%%%%%%%%%%%%%%%%%%%%%%%%%%%%%%%%%%%%%%%%%%%%%%%%%%%%%%%%%%
\subsection{Trends of age, \zh, and \afe\ with velocity dispersion}
\label{sec:trends}

\begin{figure}
\begin{center}
\leavevmode
\includegraphics[width=8cm]{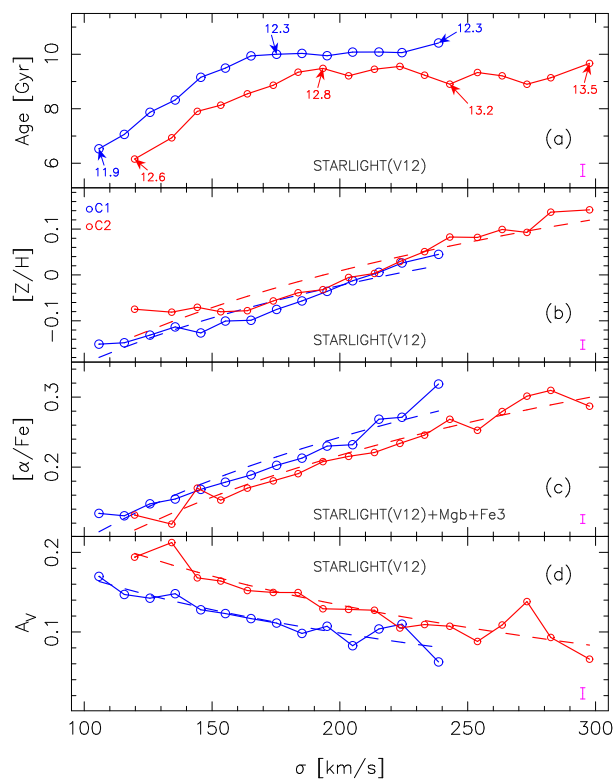}
\end{center}
\caption{The  trends  of   stellar  population  properties, $Age$ (a),
  metallicity (b), \afe\ (c), and internal reddening (d), are shown as
  a function of velocity dispersion  for central ETGs residing in low-
  (blue) and high-  (red) mass halos. The two  subsamples defined with
  respect to environment are shown with different colours as labelled in
  the  top--left  of  panel  (b).   The  magenta  error  bars  in  the
  lower--right  corner  of each  panel  show  the maximum  measurement
  uncertainty on stellar population parameters (i.e. the maximum value
  of the uncertainty among all  data-points).  The dark-grey labels in each
  panel refer to the method used to estimate the corresponding stellar
  population property. $Age$,  metallicity, and extinction are estimated
  with {\tt STARLIGHT}, while \afe\ is obtained, at fixed age, from Mg
  and Fe  spectral indices (see the  text).  In panel  (a) the average
  \lmh\ for specific bins of velocity dispersion is indicated.  }
%\contcaption{.}
\label{fig:CEN_SL}
\end{figure}

\begin{figure}
\begin{center}
\leavevmode
\includegraphics[width=8cm]{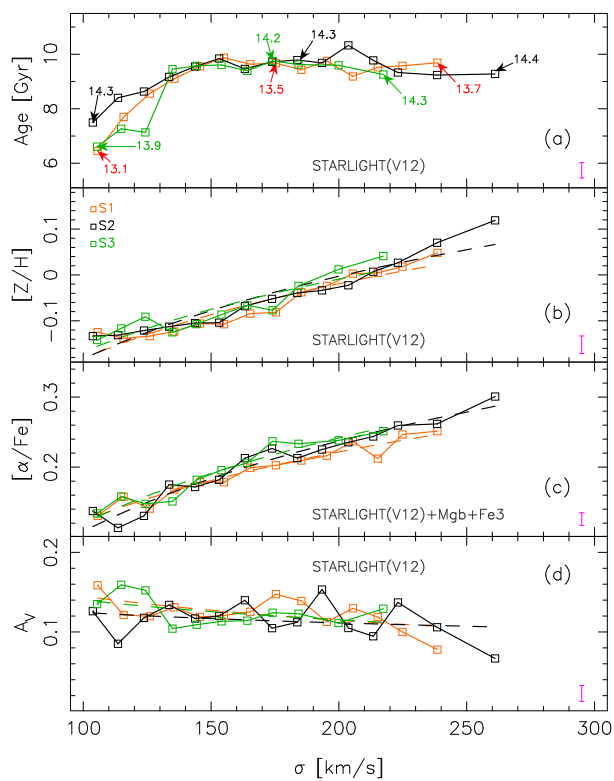}
\end{center}
\caption{Same as  Fig.~\ref{fig:CEN_SL} but for  satellite rather than
  central ETGs.  }
%\contcaption{.}
\label{fig:SAT_SL}
\end{figure}

Figs.~\ref{fig:CEN_SL} and~\ref{fig:SAT_SL}  show the main  results of
the present work, i.e. the trends  of $Age$ (panel a), \zh\ (panel b),
\afe\ (panel  c), and  \av\ (panel d)  with $\sigma$, for  central and
satellite  ETGs,  respectively,  as  a  function  of  environment.  In
App.~\ref{app:afe_ssps},      we     also     show      the     trends
(Figs.~\ref{fig:CEN_SL_CE}  and~\ref{fig:SAT_SL_CE}  for centrals  and
satellites,  respectively) but  using  $\alpha$-enhanced, rather  than
solar-scaled, MILES  models.  For \zh,  \afe, and \av, the  trends are
fitted  with   power-law  relations,  using   a  least-square  fitting
procedure with $\log \sigma$ as independent variable. For each sample,
the slopes  of the best-fit  relations, with their  uncertainties, are
reported in Tab.~\ref{tab:slopes}, the values in parentheses referring
to the case  where the stellar population parameters  are derived with
$\alpha$-MILES   (rather  than   solar-scaled   MILES)  models.    
The { intercepts}\footnote{ We notice that the { intercepts} are
  expected to be more model-dependent than the slopes, reflecting the
  uncertainties on the absolute (zero-point) calibration of stellar
  population models (with respect to both age and metallicity).  }  of
the best-fitting relations are reported in Tab.~\ref{tab:offsets}.
The uncertainties on the slopes (and { intercepts}) are obtained
from $200$ iterations, bootstrapping the residuals with respect to the
best-fitting curves.  Notice that we have not attempted to fit the
$Age$ trend as it flattens at high $\sigma$ (see below).

\begin{description}
 \item[--]  In general,  regardless of  galaxy environment,  ETGs have
   older ages, higher metallicities, and higher \afe\, with increasing
   velocity dispersion. Similar trends  are found when using $\alpha$-
   (rather      than      solar-scaled)      MILES      SSPs      (see
   App.~\ref{app:afe_ssps}).
 \item[--] The $Age$ trends become flat for $\sigma > 150$--$200$\kms.
   However, this  result is model dependent,  as for $\alpha$-enhanced
   MILES  models, the  age  keeps increasing  with  $\sigma$ also  for
   massive         ETGs         (see         Figs.~\ref{fig:CEN_SL_CE}
   and~\ref{fig:SAT_SL_CE};         and         Figs.~\ref{fig:CEN_AG}
   and~\ref{fig:SAT_AG}).
 \item[--] The  main result regarding the  dependence with environment
   is  that  the stellar  population  content  of  {\sl central}  ETGs
   depends significantly on the mass  of the halo where these galaxies
   reside.  Central ETGs in groups,  with ``high'' \mh (sample C2; see
   red       curves        in       Figures~\ref{fig:CEN_SL}       and
   Fig.~\ref{fig:CEN_SL_CE}),  have  younger  ages, with  $\delta(Age)
   \sim  -1$\,Gyr\footnote{  Differences  in  age  and  other  stellar
     population  properties  are computed  here  by interpolating  the
     trends for samples C1 and  C2 over the common $\sigma$ range from
     $130$ to  $240$\,\kms , and  taking the median difference  of each
     property.}, higher metallicities, $\delta ([Z/H]) \sim 0.02$\,dex,
   lower  \afe   ,  $\delta   $\afe  $\sim  -0.025$\,dex,   and  higher
   extinction, $\delta  $\av $\sim 0.035$~mag, than  central ETGs with
   the same velocity dispersion,  residing in lower mass halos (sample
   C1;   see   blue    curves   in   the   Fig.~\ref{fig:SAT_SL}   and
   Fig.~\ref{fig:SAT_SL_CE}).   This result  is  confirmed when  using
   $\alpha$-MILES models (Fig.~\ref{fig:CEN_SL_CE}), from the analysis
   of line strengths (Sec.~\ref{sec:ews}), or following an independent
   methodology that does not rely on stacked spectra (Gallazzi et al.,
   in   prep,   see   Appendix~\ref{app:compAG}).  The   environmental
   dependence of  the age  is detected throughout  the whole  range of
   velocity dispersion. Regarding metallicity, when using solar-scaled
   MILES models, the difference is  present at all values of $\sigma$,
   except  for two  bins with  $\sigma \sim  220$\,\kms,  whereas this
   trend is extended  to all values of velocity  dispersion when using
   the     $\alpha$-MILES      models     (see     panel      b     of
   Fig.~\ref{fig:CEN_SL_CE}).
 \item[--]  In contrast,  the  stellar populations  of satellite  ETGs
   (Fig.~\ref{fig:SAT_SL})   lack  any  significant   difference  with
   respect  to  environment, except  for  galaxies  with low  velocity
   dispersion ($\sigma < 130$\,\kms ),  where ETGs in the outskirts of
   groups (sample S3; see  green curves in Fig.~\ref{fig:SAT_SL}) have
   $\sim$1.5\,Gyr  younger ages, higher  \afe\ ($\sim  +0.02$\,dex when
   averaging  over all  three bins  with  $\sigma <  130$\,\kms ),  and
   higher  \av\  than  satellites  with  high halo  mass  (i.e.  those
   residing         in         galaxy        ``clusters'';         see
   Sec.~\ref{sec:environment}).   Regarding   the   $Age$   parameter,
   satellites in lower mass halos (sample S1; see orange curves in the
   Figures) seem to lie in between the trends found for satellite ETGs
   in the outskirts of groups and  those in more massive halos . These
   results   are   confirmed    when   using   $\alpha$-MILES   models
   (Fig.~\ref{fig:SAT_SL_CE}), as  well as  from the analysis  of line
   strengths (Sec.~\ref{sec:ews}).
 \item[--] Comparing the trends of Fig.~\ref{fig:CEN_SL} with those of
   Fig.~\ref{fig:SAT_SL}, we see that  satellite ETGs have a shallower
   $Age$--$\sigma$ relationship than central ETGs. For samples S1, S2,
   and S3, $Age$  is essentially constant for $\sigma  > 140$\,\kms. In
   contrast, the  $Age$ parameter keeps increasing up  to $\sigma \sim
   165$\kms\ ($\sim  185$\,\kms ) in centrals  at low (high)  \mh . For
   $\sigma  < 200$\,\kms,  satellites are  always older,  regardless of
   their  environment, than  groups'  centrals (sample  C1). In  other
   words, ``downsizing'' appears to be more pronounced in central ETGs
   than in  satellites (see Sec.~\ref{sec:comp_sigma}).   The internal
   extinction vs.  $\sigma$ trends are  also steeper in  the subsamples
   involving  central  ETGs  (see   the  slope  values  in  col.~4  of
   Tab.~\ref{tab:slopes}).
\end{description}

%%%%%%%%%%%%%%%%%%%%%%%%%%%%%%%%%%%%%%%%%%%%%%%%%%%%%%%%%%%%%%%%%
\subsection{Trends of line strengths with velocity dispersion}
\label{sec:ews}

To further investigate  the robustness of our results,  we analyse the
line strengths  of spectral features sensitive to  age and metallicity
in  the stacked spectra.  We consider  the age-sensitive  Balmer lines
\hbo\  and  \hgf\ ,  and  the  total  metallicity indicator  \mgfep  .
\hbo\   is  a  modified   \hb\  index,   optimised  to   minimise  the
age-metallicity  degeneracy~\citep{CV09}.  Both  \hb\  and  \hgf\  are
believed  to be  insensitive to  \afe  , in  contrast to  higher-order
Balmer lines~\citep{Thomas:04}. In order to correct the \hbo\ line for
nebular emission, we applied the same approach as in Paper VIII, where
the \hb\ spectral region (from  $4000$ to $4950$\,\AA ) is fitted with
2SSP (MILES)  models (including a  low-order multiplicative polynomial
in the fit), excluding the H$\beta$ line (from $4857$ to $4865$\,\AA )
from  the  fit. The  variation  of the  \hbo\  index  between a  given
observed  spectrum   and  its   best-fit  model  gives   the  emission
correction. The uncertainty on the emission correction is estimated as
half of the maximum variation in the emission correction when changing
the  order of  the  multiplicative polynomial  (from  $3^{\rm rd}$  to
$6^{\rm th}$ order).   We point out that the  same emission correction
approach is applied  to all stacks. No emission  correction is applied
to  \hgf, as  higher-order  Balmer lines  are  negligibly affected  by
nebular contamination in ETG spectra.  The total metallicity estimator
\mgfep\  is a  modified version  of the~\citet{GON93}  \mgfe\ spectral
index     that     removes      its     residual     dependence     on
\afe\  \citep{TMB:03}.  Hence, the  analysis  of  \hbo  , \hgf  ,  and
\mgfep\ is  virtually free  from any degeneracy  among age/metallicity
and either \afe\ or \av  , providing complementary information to that
obtained  with spectral  fitting. Since  we have  only  three spectral
indices we  do not attempt any  fitting of line  strengths with models
having varying age and metallicity, but compare instead the trends for
different samples of ETGs in a qualitative manner.

\begin{figure}
\begin{center}
\leavevmode \includegraphics[width=8.5cm]{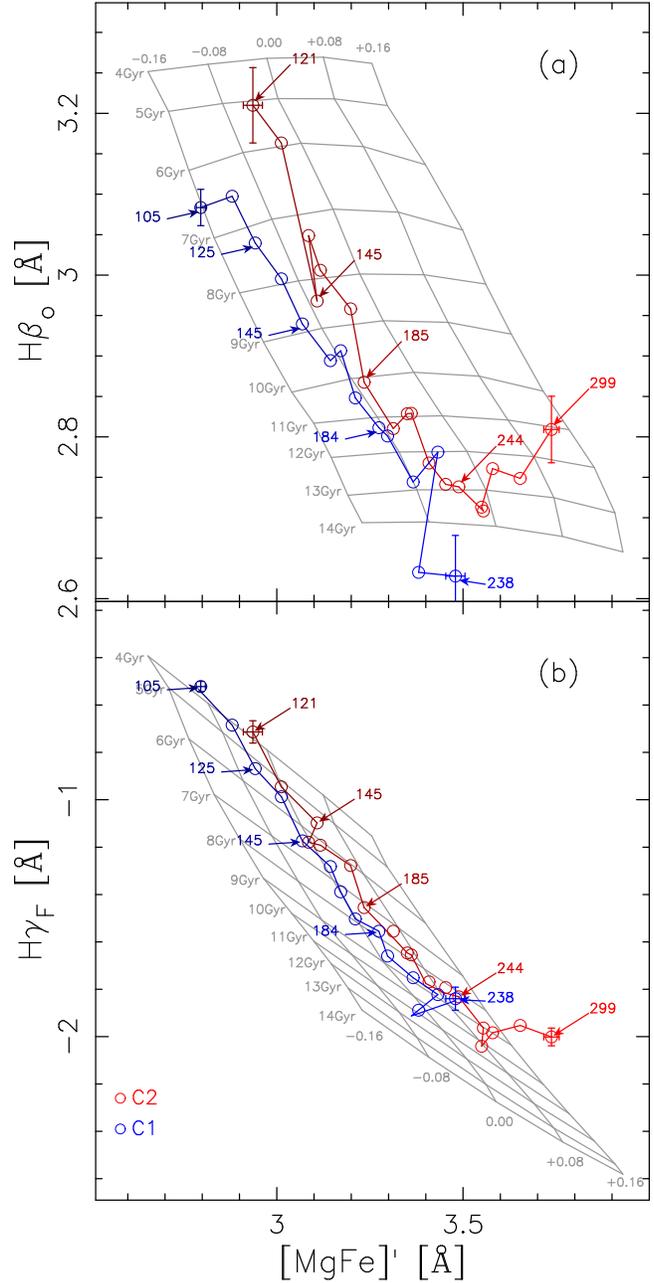}
\end{center}
\caption{The  EWs  of  the  Balmer  line  strengths  \hbo\  (top)  and
  \hgf\ (bottom) , are plotted  as a function of the total metallicity
  indicator \mgfep , for the same environment-segregated samples as in
  Fig.~\ref{fig:CEN_SL}, i.e.  central ETGs at low  and high halo-mass
  (blue and red colours, respectively).   For each sample the colour of
  the symbols and the  lines becomes lighter with increasing $\sigma$.
  Error bars (at  the 1 sigma level) are shown only  at the lowest and
  highest $\sigma$ bins, for each  sample. Arrows are used to indicate
  the $\sigma$ value  of some bins (in \kms), in order  to allow for a
  comparison of the different curves at fixed velocity dispersion. The
  grey grid  shows the  effect of varying  age and metallicity  of the
  MILES SSP models  with a Kroupa IMF. The age is  varied from 4 (top)
  to  14\,Gyr (bottom)  in steps  of 1\,Gyr,  whereas  the metallicity
  ranges  from \zh$=-0.16$  (left)  to $+0.16$  (right),  in steps  of
  $0.08$\,dex.  }
%\contcaption{.}
\label{fig:H_CEN}
\end{figure}

\begin{figure}
\begin{center}
\leavevmode
\includegraphics[width=8.5cm]{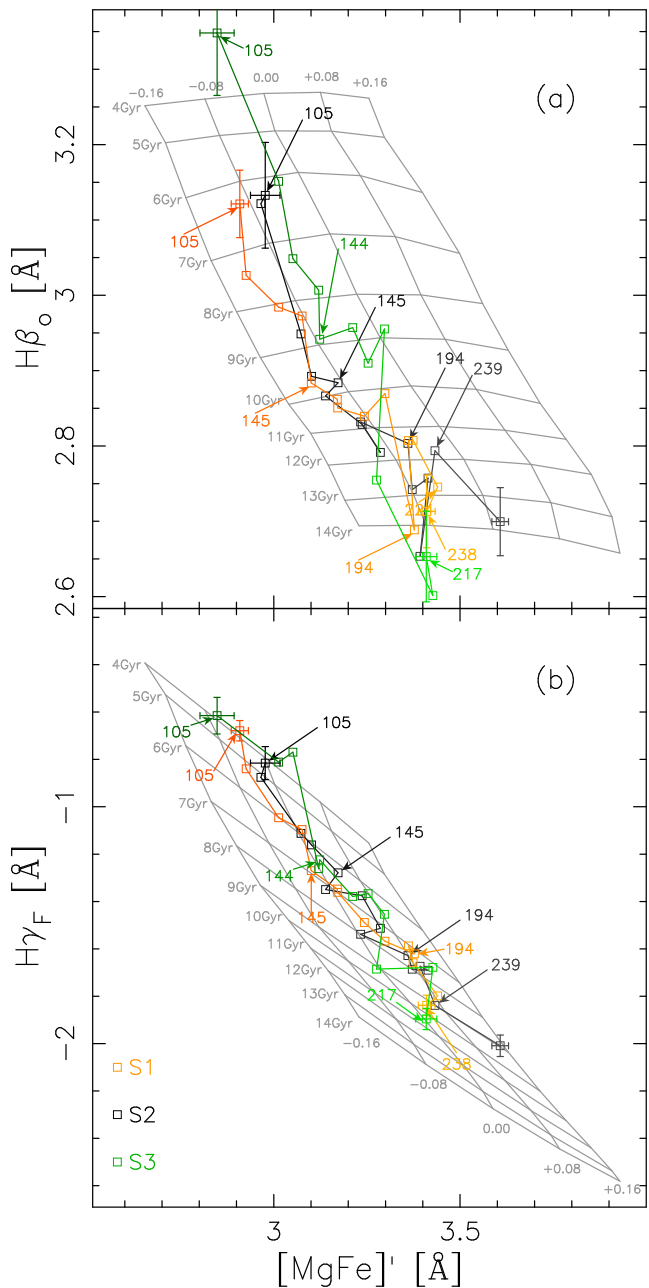}
\end{center}
\caption{ Same  as Fig.~\ref{fig:H_CEN} but for  satellite rather than
  central  ETGs.   The Figure  shows  the same  environment-segregated
  samples as in Fig.~\ref{fig:SAT_SL}.  }
%\contcaption{.}
\label{fig:H_SAT}
\end{figure}

Figs.~\ref{fig:H_CEN}  and.~\ref{fig:H_SAT}   plot  the  age-sensitive
indicators \hbo\ and \hgf\ as  a function of the metallicity estimator
\mgfep ,  for central and  satellite ETGs, respectively.   Notice that
the  relatively   large  error  bars  on  \hbo\   mainly  reflect  the
uncertainties from the emission correction, rather than from the $S/N$
ratio of  the spectra.  The  information contained in the  two Figures
can be compared with that of  the age and metallicity trends in panels
(a)   and    (b)   of   Figs.~\ref{fig:CEN_SL}   and~\ref{fig:SAT_SL},
respectively.   Colours   and   symbol   types   are   the   same   in
Fig.~\ref{fig:H_CEN}  (Fig.~\ref{fig:H_SAT}) and Fig.~\ref{fig:CEN_SL}
(Fig.~\ref{fig:SAT_SL}).  Labels  and arrows in the  Figures allow one
to  examine  the trends  in  the  line  strengths at  fixed  $\sigma$.
Fig.~\ref{fig:H_CEN} shows that central ETGs in groups (``high'' \mh )
follow  a different  sequence  than  those with  low  \mh ,  providing
further  support to the  main result  of the  present work,  i.e.  the
significant  dependence of  stellar population  properties  of central
ETGs  on  environment.   When  looking  at line  strengths  for  fixed
$\sigma$, one  can see  that at  low $\sigma$ (see  the two  bins with
$\sim 120$ and  $\sim 145$\,\kms , respectively) the  shift between the
two sequences of central ETGs (red and blue curves) is mainly due to a
``vertical''  offset, caused  by  high \mh\  centrals having  stronger
Balmer lines, i.e. younger  ages and slightly higher metallicity, than
those at low \mh . At high $\sigma$, due to the decreasing sensitivity
of Balmer lines to a variation of age for old stellar populations, the
origin of the difference between the two sequences becomes less clear.
These results are qualitatively consistent with the offset between red
and        blue       curves        in       panel        (a)       of
Fig.~\ref{fig:CEN_SL}.  Fig.~\ref{fig:H_SAT} shows that  all satellite
samples  exhibit similar trends  in the  \hbo --  \mgfep\ and  \hgf --
\mgfep\  diagrams,  with  a  tendency  for  satellites  in  the  group
outskirts (green curves)  to have younger ages, at  low $\sigma$, with
respect to  the other satellite  samples, consistent with  the results
from spectral fitting (Fig.~\ref{fig:SAT_SL}).

\section{Possible systematic effects}
\label{sec:systematics}

\begin{figure}
\begin{center}
\leavevmode
\includegraphics[width=8cm]{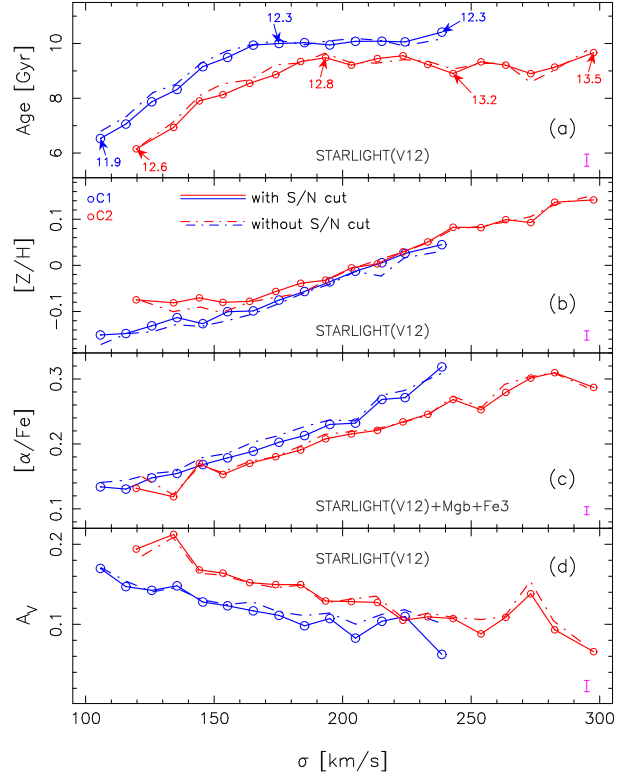}
\end{center}
\caption{Comparison of trends of stellar population properties of ETGs
  obtained with (solid) and without (dot-dashed) selecting SDSS
  spectra with best $S/N$ ratio (see the text). Notice that solid
  curves and labels are the same as in Fig.~\ref{fig:CEN_SL}.}
%\contcaption{.}
\label{fig:CEN_SL_NOSN}
\end{figure}

\begin{figure}
\begin{center}
\leavevmode
\includegraphics[width=8cm]{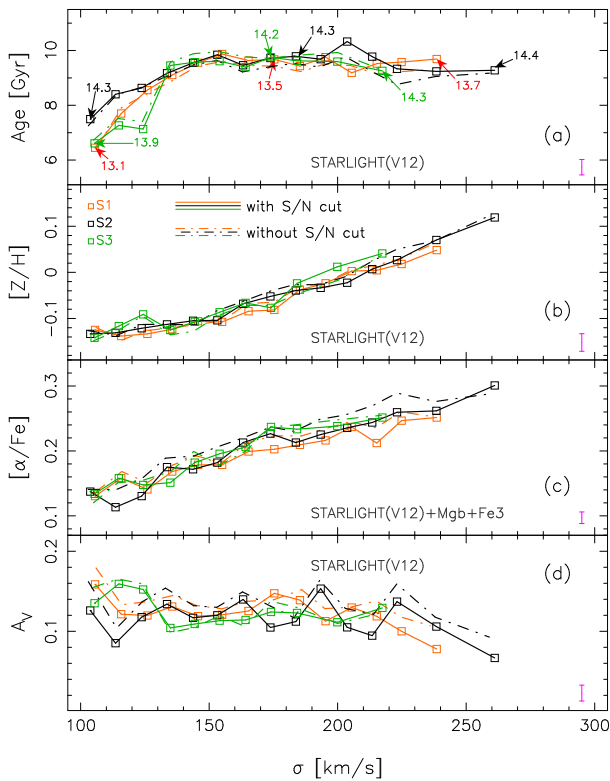}
\end{center}
\caption{Same as Fig.~\ref{fig:CEN_SL_NOSN} but for the samples of satellite ETGs. Solid curves and labels are the 
same as in Fig.~\ref{fig:SAT_SL}.}
%\contcaption{.}
\label{fig:SAT_SL_NOSN}
\end{figure}

\begin{figure}
\begin{center}
\leavevmode
\includegraphics[width=8cm]{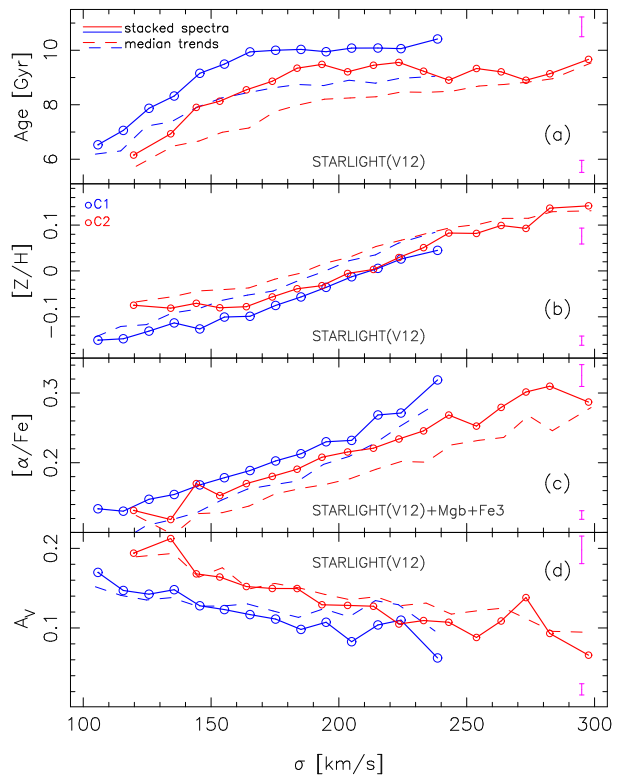}
\end{center}
\caption{ Comparison of the trends of stellar population properties
  of central ETGs obtained from our stacked spectra (solid curves) and
  by taking median values of the estimates for {\sl individual} SDSS
  spectra in each velocity dispersion bin (dashed curves). Notice that
  solid curves and the magenta error bars in the lower--right of each
  panel are the same as in Fig.~\ref{fig:CEN_SL}. Magenta error bars
  are maximum uncertainties (among bins and samples) for
  median-combined (dashed curves) trends.  }
%\contcaption{.}
\label{fig:CEN_STACKS_MEDIAN}
\end{figure}

{ 

We present in this section a battery of tests proving that the results
shown in Figs.~\ref{fig:CEN_SL} and~\ref{fig:SAT_SL} are not affected
by systematics. These tests can be summarized as follows:

\begin{description}
 \item[{\it Correlated uncertainties - }] An important issue for any
   study of stellar populations is the effect of degeneracies among
   the various properties that define a population.  In particular,
   underestimating the age of a stellar population can be balanced by
   an increase of its estimated metallicity~\citep{wo94} and dust
   content.  We notice that the results of the present work, and in
   particular the difference between the trends of samples C1 and C2,
   are not affected by this degeneracy. Firstly, there is no reason
   why such a degeneracy should affect only the samples of centrals,
   and not that of satellite ETGs. More importantly, differences
   between samples C1 and C2 are confirmed by an independent analysis
   of line strengths (see below, and Appendix~\ref{app:compAG}), which
   is unaffected by the age--dust degeneracy, and relatively
   unaffected by the age--metallicity degeneracy (when relying on the
   \hbo\ age indicator, see Sec.~\ref{sec:ews}).  In
   App.~\ref{app:afe_deg}, we also show that measurement errors tend
   to shift the stellar population properties of samples C1 and C2
   along parallel loci in the space of $Age$, \zh , and \av
   . Therefore, the differences between the blue and red curves in
   each panel of Fig.~\ref{fig:CEN_SL} are not artefacts from stellar
   population degeneracies or caused by a particular marginalisation of the
   uncertainties along a specific direction in the parameter space
   explored.

\item[{\it Uncertainties on \mh\ - }] We have also verified that the
  stellar population trends for central ETGs are not affected by the
  uncertainties on halo mass estimates.  To this effect, we have
  repeated the analysis by re-defining samples C1 and C2 with
  different mass thresholds of \lmha$<12.3$ and \lmhb$>12.7$,
  respectively, rather than using a single separation value of
  \lmh$=12.5$ (see Sec.~\ref{sec:environment}).  The difference
  between \mh$_{,1}$ and \mh$_{,2}$ is roughly twice the group mass
  uncertainty estimated by Y07 for \lmh$\sim 12$ (see
  Sec.~\ref{sec:environment}).  We found median differences of
  $\delta(Age) \sim -1.1$\,Gyr, $\delta ([Z/H]) \sim 0.02$\,dex, $\delta
  $\afe $\sim -0.038$\,dex, and $\delta $\av $\sim 0.048$\,mag, between
  group and isolated centrals, consistent with
  Fig.~\ref{fig:CEN_SL}.

\item[{\it S/N ratio of SDSS spectra - }] Our reference stacks are
  obtained by excluding spectra whose $S/N$ ratio is below the lowest
  quartile of the $S/N$ distribution of all ETG spectra within each
  $\sigma$ bin (see Sec.~\ref{sec:sample}). We have repeated the
  analysis by removing this cut in $S/N$. The size of the sample
  increases from $N_{ETGs}=20,977$ to $28,356$. The fraction of ETGs
  excluded by the $S/N$ cut amounts to $f_{\rm S/N} \sim 28 \%$, $12
  \%$, $27 \%$, $32 \%$, and $30 \%$, for samples C1, C2, S1, S2, and
  S3, respectively. Notice that the lowest value of $f_{\rm S/N}$
  corresponds to sample C2 as galaxies in the more massive halos are
  brighter (see below). Although $f_{\rm S/N} $ varies among different
  samples, the results of Figs.~\ref{fig:CEN_SL} and~\ref{fig:SAT_SL}
  turn out to be unaffected by the $S/N$ selection. This is shown in
  Figs.~\ref{fig:CEN_SL_NOSN} and~\ref{fig:SAT_SL_NOSN}, where we
  compare the trends of stellar population parameters among stacked
  spectra obtained with (solid curves) and without (dot-dashed curves)
  the $S/N$ selection. Relative differences among samples (e.g. C1 and
  C2, as well as those among S1, S2, and S3 at lowest $\sigma$) are
  completely unaffected by the $S/N$ cut.

\item[{\it Trends for individual and stacked spectra - }] As mentioned
  in Sec.~\ref{sec:trends} and shown in Appendix~B, the trends of
  Figs.~\ref{fig:CEN_SL} and~\ref{fig:SAT_SL} are confirmed when using
  an independent methodology that relies on individual, rather than
  stacked, spectra, as well as a different set of stellar population
  models (BC03 rather than MILES), to infer the $Age$, \zh , and
  \afe\ parameters (Gallazzi et al., in prep). We have also derived
  $Age$, \zh , \afe , and \av\ from individual spectra with the same
  identical approach as for stacked spectra (Secs.~\ref{sec:age_zh}
  and~\ref{sec:alpha}).  Fig.~\ref{fig:CEN_STACKS_MEDIAN} compares the
  trends with $\sigma$ for central ETGs\footnote{The same comparison
    for satellites is not shown for brevity reasons, as it shows the
    same features as in Fig.~\ref{fig:CEN_STACKS_MEDIAN}.} when using
  stacked spectra (solid curves, i.e. the same as in
  Fig.~\ref{fig:CEN_SL}) and median-combining (dashed curves) the
  estimates from individual spectra.  Although we have run STARLIGHT with the same 
  setup, and in particular the same input basis of SSP models (see
  Sec.~\ref{sec:sp_pars}), for both stacked and individual spectra, some differences 
  exist between
  the two approaches, as age and \afe\ are systematically lower (by
  $\sim 2$\,Gyr and $\sim 0.04$\,dex, respectively), while \zh\ is
  systematically higher (by $\sim 0.04$\,dex) for median values with
  respect to stacked spectra. 
%  \textcolor{blue}
  {The fact that the median ages from individual spectra are lower than the estimates
  from stacked spectra is likely due to the larger number~\footnote{
  More in detail, the STARLIGHT input SSPs span 
  27 age values (Sec.~\ref{sec:sp_pars}). Out of them, 20 (7) ages are younger (older) 
  than 8~Gyr.} of young
  ($\simlt$8--9\,Gyr, typical for ETGs in our sample), relative to old (i.e. $\simgt$8--9\,Gyr), SSPs in the
  input basis provided to STARLIGHT}. Because of that, at the relatively
  low-$S/N$ ratio of individual SDSS spectra, young SSPs get a
  non-zero weight more frequently than the old ones in the
  best-fitting STARLIGHT mixture, hence leading to lower age values
  (and thus higher \zh, because of the age--metallicity degeneracy)
  when averaging over a given set of spectra. Nevertheless, as shown
  in Fig.~\ref{fig:CEN_STACKS_MEDIAN}, {\it relative} differences
  among samples still hold, regardless of the approach adopted.
%  \textcolor{red}{Francesco, this last point is rather confusing. It
%  reads like you have decided to do two different STARLIGHT fits,
%  with the one for the individual spectra using younger SSPs in the
%  basis. It would be much more consistent if you used the same basis 
%  in all cases: individual and stacked, and then find -- if you do --
%  that lower S/N will give more weight to the younger SSPs.}

\item[{\it Selection effects - }] Figs.~\ref{fig:CEN_MR}
  and~\ref{fig:SAT_MR} show the median trends (solid curves) of
  stellar mass, \mstar , and effective radius, \rad , as a function of
  $\sigma$ for central and satellite ETGs, respectively, with dashed
  lines marking the 16th and 84th percentiles of the
  distributions. Stellar masses have been estimated from the available
  photometry (SDSS griz plus UKIDSS YJHK photometry) for each SPIDER
  ETG, as detailed in Paper V, while effective radii have been
  obtained by processing SDSS frames with the software 2DPHOT~(see
  Paper I for details). While no significant difference is seen among
  the three samples of satellites, samples C1 and C2 differ
  significantly, with ETGs at high halo-mass having larger \rad\ and
  \mstar\ than those at low \mh . This difference is because of the
  correlation between \mstar\ and \mh\ for centrals, and the
  mass--size relation of ETGs~(see, e.g.,~\citealt{Bernardi:2014} and
  references therein).  Since age is generally found to increase with
  galaxy mass~(e.g.~\citealt{Gallazzi:06}), and ETGs have shallow age
  gradients in their inner regions~\citep{LaBarbera:12}, any
  difference in \rad\ and \mstar\ between C1 and C2 should not affect
  those presented in Fig.~\ref{fig:CEN_SL} (at least the age--$\sigma$
  trends in panel a). To address this issue more explicitly, i.e. to
  test whether differences in the stellar population properties of
  centrals are affected by those in mass and radius, we select two
  subsamples of ETGs from C1 and C2, respectively, consisting of
  galaxies within the {\it same} range of \mstar\ and \rad\ for
  both samples. In practice, we select C1 and C2 ETGs within the 16th
  and 84th percentiles of the \mstar\ and \rad\ distributions for
  C1\footnote{For $\sigma>240$\,\kms , where there are no bins for C1,
    we only select ETGs from C2 with \rad$<7$\,kpc, corresponding to
    the 84th percentile of the \rad\ distribution for C1 at $\sigma
    \sim 240$\,\kms.  }  (i.e. C1 and C2 galaxies that lie within the
  blue-dashed curves in both the upper and lower panels of
  Fig.~\ref{fig:CEN_MR}).  This selection produces two samples of
  $5,550$ (C1) and $795$ (C2) ETGs, whose median \mstar\ and
  \rad\ differ by less than 0.1\,dex for all $\sigma$ bins (see black
  and magenta dot-dashed curves in Fig.~\ref{fig:CEN_MR}).  Hereafter,
  we refer to these subsamples as C1$($\rad$;$\mstar$)$ and
  C2$($\rad$;$\mstar$)$, respectively.  Fig.~\ref{fig:CEN_SL_MATCHED}
  compares the median trends of stellar population properties for C1
  and C2 (solid curves; the same as in
  Fig.~\ref{fig:CEN_STACKS_MEDIAN}) to those for C1$($\rad$;$\mstar$)$
  and C2$($\rad$;$\mstar$)$ (dot-dashed curves).  Remarkably, although the \mstar\ and
  \rad\ selection reduces the size of the parent samples C1 and C2,
  the trends of stellar population properties do not change
  significantly, confirming the robustness of our result, i.e. that
  central ETGs at high \mh\ have younger ages, lower
  $\alpha$-enhancement, higher extinction (and slightly higher \zh)
  than those residing in lower mass halos.

\item[{\it Aperture  effects - }]  SDSS spectra are observed  within a
  fixed aperture of 3$''$ diameter. Because of the existence of radial
  gradients of  stellar population  properties in ETGs,  the use  of a
  fixed aperture  might bias  stellar population trends  with $\sigma$
  and   environment.    Indeed,   as    found   in   Paper   II   (see
  also~\citealt{Scodeggio:98}),  trends of  age  and metallicity  with
  galaxy mass can be significantly shallower when these parameters are
  estimated for a  whole galaxy, rather than its  central regions. For
  the purpose of the present  work, we assess whether aperture effects
  can  introduce differences  among samples  of central  and satellite
  ETGs.  Tab.~\ref{tab:fiber_diam} reports  the median fibre diameter,
  \dfib  ,  and  the  lowest  and  highest  16th  percentiles  of  the
  \dfib\ distributions,  for the  samples of centrals  and satellites.
  The values  of \dfib\ have  been computed after rescaling,  for each
  galaxy, the fibre diameter  to physical projected distances (i.e. in
  kpc units), using SDSS  spectroscopic redshifts. The median value of
  \dfib\ is  consistent among samples,  amounting to $\sim  4$\,kpc in
  general.   Comparing  the   fibre  projected  physical  distance  to
  effective  radii,  we  find  for  satellite ETGs  that  the  typical
  \rad\  is  $\sim  4$\,kpc  in  all  samples (S1,  S2,  and  S3;  see
  Fig.~\ref{fig:SAT_MR}), meaning that the SDSS fibre samples the same
  galactocentric distance  of about \rad/2 for  all satellites. Hence,
  aperture effects do not introduce a bias in the comparison of trends
  for S1, S2,  and S3 in Fig.~\ref{fig:SAT_SL}.  For  central ETGs, we
  notice that while the median \dfib\  is the same for C1 and C2, ETGs
  in C2 have larger \rad\ (by $\sim 0.3$\,dex, i.e. a factor of two in
  linear units)  than those in C1  (Fig.~\ref{fig:CEN_MR}). Hence, the
  SDSS  fibre  samples   a  more  central  region  for   C2  (i.e.   a
  galactocentric  distance of  $\sim$\rad$/4$)  than for  C1 (i.e.   a
  galactocentric  distance of  $\sim$\rad$/2$). {  Since  ETGs have
    negative  metallicity gradients (i.e.   higher metallicity  in the
    centre  than in the  outskirts; e.g.~\citealt{Peletier:1990}) },
  one could  expect that aperture  effects may be responsible  for the
  difference   in    \zh\   between   C1   and   C2    (panel   b   of
  Fig.~\ref{fig:CEN_SL}).  However, this  is not necessarily the case,
  as  the seeing,  whose typical  extent  for SDSS  data is  FWHM$\sim
  1.4''$ in the r band, tends to wash out stellar population gradients
  over a  spatial scale  of a few  arcsec (mapping  into a scale  of $
  \sim$\rad$/2$  in  C1).   Tab.~\ref{tab:fiber_diam} shows  that  the
  subsamples C1$($\rad$;$\mstar$)$ and C2$($\rad$;$\mstar$)$, selected
  to span  the same range of  \rad\ and \mstar\ (see  above), have the
  same range of  \dfib\ (i.e. the same median  and percentile values),
  i.e.   the SDSS  fibre spans  the same  galactocentric  distance, in
  units    of   \rad    ,   for    both    C1$($\rad$;$\mstar$)$   and
  C2$($\rad$;$\mstar$)$.  Since stellar population differences between
  C1$($\rad$;$\mstar$)$  and  C2$($\rad$;$\mstar$)$  are the  same  as
  those between  C1 and C2, we  conclude that aperture  effects do not
  affect the results of our work.
\end{description}

}

\begin{table}
\centering
\small
\begin{minipage}{80mm}
\caption{ Median values, among different $\sigma$ bins, of the 16th
  (col.~2), 50th (i.e. the median; col.~3), and 84th (col.4)
  percentiles of (3$''$) SDSS-fibre diameter values, rescaled to kpc
  according to galaxy spectroscopic redshifts, for different samples
  of ETGs. Values in parentheses are standard deviations (among
  $\sigma$ bins).  Samples C1(\rad; \mstar ) and C2(\rad; \mstar ) are
  obtained from C1 and C2 by selecting ETGs in the same range of
  \mstar\ and \rad (see the text).  }
  \begin{tabular}{c|c|c|c}
 \hline
  SAMPLE & \multicolumn{3}{c}{fibre diameter (kpc)} \\
          & 16th percentile & median & 84th percentile \\
      (1) & (2) & (3) & (4) \\
 \hline
C1 &                 $3.48 (0.14)$ & $4.13 (0.20)$ &  $4.82 (0.16)$ \\
C2 &                 $3.50 (0.15)$ & $4.23 (0.26)$ &  $4.93 (0.19)$ \\
S1 &                 $3.46 (0.11)$ & $4.14 (0.15)$ &  $4.83 (0.07)$ \\
S2 &                 $3.48 (0.11)$ & $4.10 (0.14)$ &  $4.78 (0.08)$ \\
S3 &                 $3.44 (0.14)$ & $4.05 (0.13)$ &  $4.72 (0.08)$ \\
C1(\rad; \mstar ) &  $3.49 (0.07)$ & $4.10 (0.15)$ &  $4.85 (0.15)$ \\
C2(\rad; \mstar ) &  $3.42 (0.20)$ & $4.05 (0.28)$ &  $4.92 (0.23)$ \\
\hline
\end{tabular}
\label{tab:fiber_diam}
\end{minipage}
\end{table}

\begin{figure}
\begin{center}
\leavevmode
\includegraphics[width=8cm]{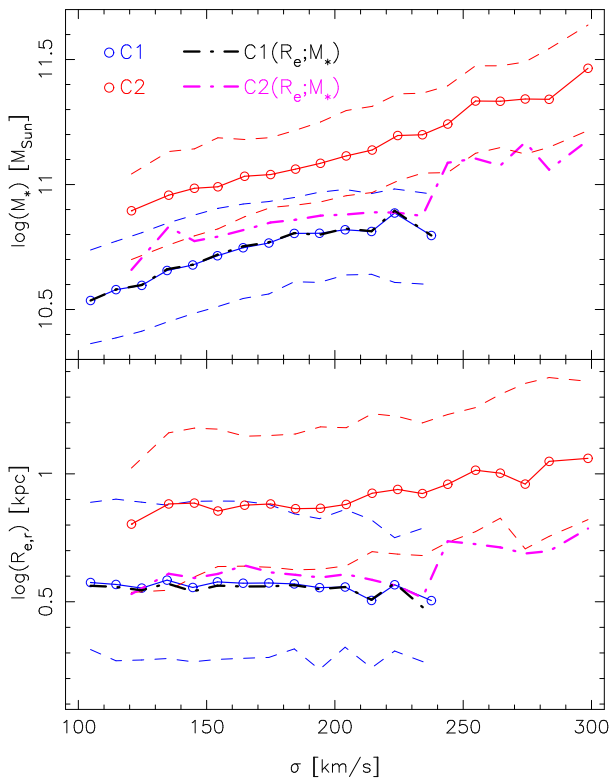}
\end{center}
\caption{ Trends of logarithmic stellar mass (top) and (r-band)
  effective radius (bottom) as a function of velocity dispersion for
  central ETGs residing in low-(blue) and high-(red) mass halos
  (i.e. samples C1 and C2, respectively; see top--left labels).  Solid
  curves are median trends, while dashed curves mark the 16th-84th
  percentiles of the distributions for each $\sigma$ bin.
  Because of the \mstar--halo mass correlation for centrals and the
  mass-size relation of ETGs, galaxies in sample C2 tend to be more
  massive and larger than those in C1. 
  Black and magenta dot-dashed curves show the trends for the subsamples 
  C1$($\rad$;$\mstar$)$ and C2$($\rad$;$\mstar$)$, obtained by selecting ETGs
  with the same range of \rad\ and \mstar\ from samples C1 and C2, respectively.
  %  \textcolor{red}{The cyan line may be hard to see. Use other colour?
%  Black should be fine.}
 }
%\contcaption{.}
\label{fig:CEN_MR}
\end{figure}

\begin{figure}
\begin{center}
\leavevmode
\includegraphics[width=8cm]{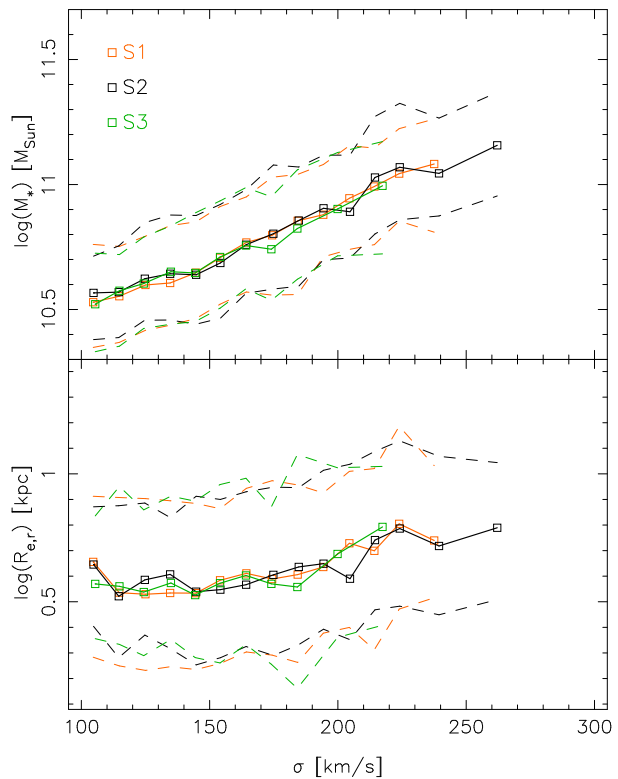}
\end{center}
\caption{  Same as Fig.~\ref{fig:CEN_MR} but for the three samples
  of satellite ETGs.  Notice that the distributions of stellar mass
  and effective radius are now independent of the
  environment.  }
%\contcaption{.}
\label{fig:SAT_MR}
\end{figure}

\begin{figure}
\begin{center}
\leavevmode
\includegraphics[width=8cm]{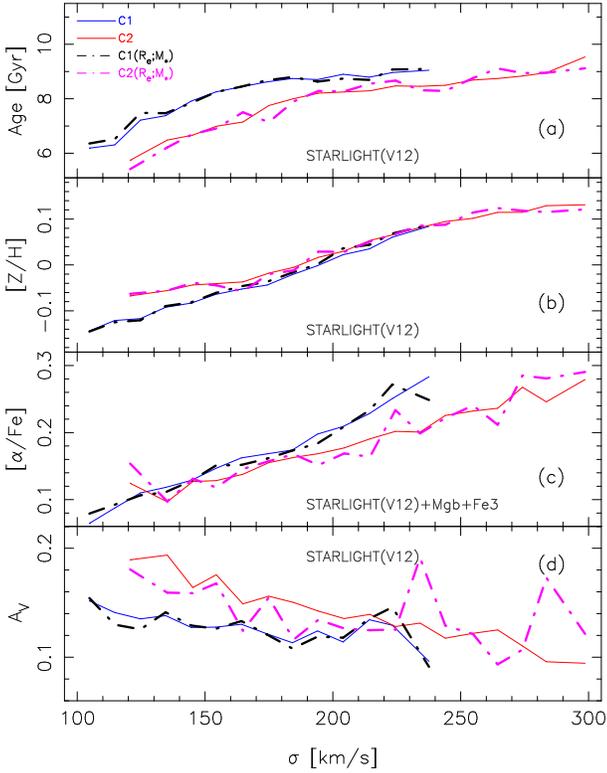}
\end{center}
\caption{ Comparison of median trends of stellar population
  properties for samples C1 and C2 (solid curves), with those  
  (dot-dashed curves) for subsamples C1$($\rad$;$\mstar$)$ and
  C2$($\rad$;$\mstar$)$ (plotted as dot-dashed black and magenta curves
  also in Fig.~\ref{fig:CEN_MR}). These subsamples are constructed from C1
  and C2 to span the same range of stellar mass and galaxy size
  (see the text for details). Notice that the solid curves are the
  same as in Fig.~\ref{fig:CEN_STACKS_MEDIAN} for reference.
%  \textcolor{red}{Just in case, the grey labels may be too
%    faint. Sometimes it is not possible to see this type of subtle
%    colouring in printed versions. I would make it black, or at least
%    dark grey. Likewise in the rest of figures where this colouring is
%    shown. Your call though.} 
}
%%\contcaption{.}
\label{fig:CEN_SL_MATCHED}
\end{figure}

%%%%%%%%%%%%%%%%%%%%%%%%%%%%%%%%%%%%%%%%%%%%%%%%%%%%%%%%%%%%%%%%%
\section{Comparison to previous works}
\label{sec:comps}

%%%%%%%%%%%%%%%%%%%%%%%%%%%%%%%%%%%%%%%%%%%%%%%%%%%%%%%%%%%%%%%%%
\subsection{Trends with velocity dispersion}
\label{sec:comp_sigma}
Since the  slopes of \zh\ and  \afe\ with $\sigma$  are independent of
environment  (Tab.~\ref{tab:slopes}),  it  is  useful  to  derive  the
average slopes among all five samples of ETGs, i.e.
\begin{equation}
\beta_{Z \! / \! H}=\frac{\delta([Z/H])}{\delta(\log \sigma)}=0.58 \pm 0.05,
\end{equation}
and 
\begin{equation}
\beta_{\alpha \! / \! {\rm Fe}}=\frac{\delta([\alpha/{\rm Fe}])}{\delta(\log \sigma)}=0.42 \pm 0.06. 
\end{equation}
These  values   can  be  compared   with  those  obtained   from  past
studies~\citep{Trager:00,  Bernardi:03,  Thomas:05, Smith:07,  Zhu:10,
  Thomas:10,   Harrison:11}.     As   shown   in    \citet[see   their
  table~5]{Harrison:11},  different  studies  found  a wide  range  of
values of \betazh\ (\betaafe), from $\sim 0.18$ ($0.2$) to $\sim 0.79$
($0.36$).  This  range of  values is likely  due to  different methods
used  to derive stellar  population parameters,  as well  as different
selection  criteria  of  the  samples.   Our values  of  \betazh\  and
\betaafe\ overlap, within the  uncertainties, with the range of values
in the  literature. In  particular, despite the  different methodology
used  to  derive   \zh\  and  \afe  ,  our   values  of  \betazh\  and
\betaafe\  are fairly  consistent,  at the  1.5\,$\sigma$ level,  with
those recently obtained  by~\citet{Thomas:10}, i.e.  \betazh$=0.65 \pm
0.02$ and \betaafe$=0.33 \pm 0.01$, for morphological-selected samples
of  ETGs drawn  from the  SDSS.  Regarding  the $Age$  parameter, most
studies agree, in particular at  low $\sigma$, that age increases with
velocity  dispersion.   This finding  fits  well  with the  downsizing
picture of  galaxy formation, where  the typical mass of  star forming
galaxies is  seen to shift  towards higher values with  look-back time
(e.g.~\citealt{Cowie:96}).   We notice that  the increasing  trends of
$Age$  and  \zh\  with  $\sigma$  also agree  qualitatively  with  our
previous work~\citep[PGF10]{Pasquali:10}, where we analysed the trends
of such parameters with  stellar mass (instead of velocity dispersion)
for the whole population of galaxies (rather than selecting ETGs only,
as in the present  work), as well as with the trends  as a function of
stellar  and   dynamical  mass  found   for  ETGs  (selected   by  the
concentration  of  the   light  profile)  in~\citet{Gallazzi:06}.   We
emphasise  that  although  the  trends  presented  in  this  work  are
consistent with previous works, our study allows us to explore for the
first time  how at fixed  velocity dispersion, the  stellar population
parameters  depend in a  different way  on halo  mass for  central and
satellite ETGs.

%%%%%%%%%%%%%%%%%%%%%%%%%%%%%%%%%%%%%%%%%%%%%%%%%%%%%%%%%%%%%%%%%
\begin{table*}
\centering
\small
\begin{minipage}{130mm}
 \caption{Slopes   of   the   correlations  between   \zh\   (col.~2),
   \afe\ (col.~3) and  \av\ (col.~4), vs. \ls ,  for different samples
   of ETGs (col.~1; see Tab.~\ref{tab:samples}).  Errors are quoted at
   the  1\,$\sigma$ level.  Values in  parentheses are  those obtained
   with $\alpha$-enhanced (rather than ``standard'') MILES models.  }
  \begin{tabular}{c|c|c|c}
 \hline
  SAMPLE & \zh\ slope & \afe\ slope & \av\ slope \\
    (1) & (2) & (3) & (4) \\
 \hline
 $\rm C1$ & $ 0.57 \pm 0.07 (0.65 \pm 0.03) $ & $ 0.48 \pm 0.06 (0.49 \pm 0.06) $ & $-0.23 \pm 0.03 (-0.32 \pm 0.03) $ \\ 
 $\rm C2$ & $ 0.64 \pm 0.07 (0.66 \pm 0.04) $ & $ 0.47 \pm 0.04 (0.44 \pm 0.04) $ & $-0.29 \pm 0.04 (-0.34 \pm 0.03) $ \\
 $\rm S1$ & $ 0.54 \pm 0.07 (0.62 \pm 0.05) $ & $ 0.33 \pm 0.03 (0.33 \pm 0.03) $ & $-0.10 \pm 0.05 (-0.15 \pm 0.05) $ \\
 $\rm S2$ & $ 0.60 \pm 0.06 (0.59 \pm 0.04) $ & $ 0.43 \pm 0.03 (0.41 \pm 0.03) $ & $-0.04 \pm 0.06 (-0.12 \pm 0.05) $ \\
 $\rm S3$ & $ 0.53 \pm 0.08 (0.61 \pm 0.05) $ & $ 0.40 \pm 0.03 (0.41 \pm 0.04) $ & $-0.08 \pm 0.07 (-0.16 \pm 0.04) $ \\
 \hline
\end{tabular}
\label{tab:slopes}
\end{minipage}
\end{table*}

\begin{table*}
\centering
\small
\begin{minipage}{140mm}
 \caption{{ Intercepts } of  the   correlations   between  \zh\   (col.~2),
   \afe\ (col.~3) and  \av\ (col.~4), vs. \ls ,  for different samples
   of  ETGs  (col.~1; see  Tab.~\ref{tab:samples}).   The { intercepts}  are
   computed at a reference  velocity dispersion of $200$\,\kms . Errors
   are  quoted at  the 1\,$\sigma$  level. Values  in  parentheses are
   those  obtained with  $\alpha$-enhanced (rather  than ``standard'')
   MILES models.  }
  \begin{tabular}{c|c|c|c}
 \hline
  SAMPLE & \zh\ intercept & \afe\ intercept & \av\ intercept \\
    (1) & (2) & (3) & (4) \\
 \hline
 $\rm C1$ & $ -0.023 \pm 0.006 (0.015 \pm 0.003) $ & $ 0.243 \pm 0.006 (0.247 \pm 0.006) $ & $0.099 \pm 0.005 (0.059 \pm 0.005) $ \\ 
 $\rm C2$ & $  0.009 \pm 0.008 (0.046 \pm 0.003) $ & $ 0.217 \pm 0.003 (0.224 \pm 0.003) $ & $0.134 \pm 0.003 (0.097 \pm 0.003) $ \\
 $\rm S1$ & $ -0.019 \pm 0.006 (0.032 \pm 0.004) $ & $ 0.222 \pm 0.003 (0.229 \pm 0.004) $ & $0.115 \pm 0.006 (0.083 \pm 0.006) $ \\
 $\rm S2$ & $ -0.003 \pm 0.006 (0.042 \pm 0.005) $ & $ 0.237 \pm 0.003 (0.241 \pm 0.003) $ & $0.111 \pm 0.007 (0.073 \pm 0.009) $ \\
 $\rm S3$ & $ -0.009 \pm 0.012 (0.039 \pm 0.008) $ & $ 0.242 \pm 0.005 (0.246 \pm 0.006) $ & $0.115 \pm 0.007 (0.076 \pm 0.006) $ \\
 \hline
\end{tabular}
\label{tab:offsets}
\end{minipage}
\end{table*}

% $C1$ & $ 0.57 \pm 0.07 (0.84 \pm 0.06)$ & 
% $C2$ & $ 0.64 \pm 0.07 (0.84 \pm 0.04)$ & 
% $S1$ & $ 0.54 \pm 0.07 (0.83 \pm 0.04)$ & 
% $S2$ & $ 0.60 \pm 0.06 (0.89 \pm 0.10)$ & 
% $S3$ & $ 0.53 \pm 0.08 (0.90 \pm 0.15)$ & 
%%%%%%%%%%%%%%%%%%%%%%%%%%%%%%%%%%%%%%%%%%%%%%%%%%%%%%%%%%%%%%%%%

%%%%%%%%%%%%%%%%%%%%%%%%%%%%%%%%%%%%%%%%%%%%%%%%%%%%%%%%%%%%%%%%%
\subsection{Trends with environment}
\label{sec:comp_env}

As reported in Sec.~\ref{sec:intro}, most of the previous studies have
analysed  stellar  population properties  of  ETGs  as  a function  of
``local''  environment,  characterised  either  through  local  galaxy
density  or   by  splitting  ETGs  into  field   and  cluster  samples
(e.g.~\citealt{Bernardi:03,     BERN:06,     Thomas:05,     Thomas:10,
  Annibali:07, Clemens:09,  Zhu:10, Cooper:10}).  A  general consensus
has emerged that, at fixed galaxy mass, either the whole population or
some  fraction of  ETGs in  low-density regions  host  younger stellar
populations than  those in dense  regions (i.e.  cluster  cores). This
difference is found to increase towards lower mass galaxies.  In order
to compare our findings with these works, we use the definition of ETG
environment as  in ~\citet[hereafter Paper  III]{SpiderIII}.  Using an
updated  FoF  group catalogue  created  as in~\citet{Berlind:06}  (the
difference  between the  two being  in  the area  used, i.e.  SDSS/DR7
rather than  DR3), in Paper III  we classified SPIDER  ETGs into group
members,  field  galaxies  (i.e.   objects with  no  group  membership
assigned, and those far from any FoF group), and un-classified objects
(mostly  galaxies  residing  in   poor  groups  and/or  in  the  group
``borders'',  see  Paper  III   for  details).   These  three  classes
constitute  46\%,  33\%,  and   21\%  of  the  entire  SPIDER  sample,
respectively.

We note that  the present study relies on  a different group catalogue
than  that used  in Paper  III  (i.e.  the  Yang et  al., rather  than
Berlind  et  al.,  catalogue).   Although  a  detailed  comparison  of
different  group  catalogues is  certainly  beyond  the  scope of  the
present work, it is instructive to analyse the composition of the five
samples of central  and satellite ETGs in terms  of the environment as
defined  in Paper  III. Tab.~\ref{tab:field_group}  reports,  for each
sample, the  fraction of field, group, and  un-classified objects.  As
expected,  the three  samples  of  satellites (S1,  S2,  and S3)  only
include a negligible fraction of field galaxies, especially sample S2,
i.e.   galaxies residing  in  the central  regions  of massive  groups
(``clusters''), where  no galaxy is classified  as ``field'' according
to Paper III.  The Table  also reveals that the population of centrals
is  much  more  heterogeneous,  including a  significant  contribution
($>30$\%)   from  field   galaxies,  in   both  samples   C1   and  C2
(i.e. regardless of halo mass),  { with a larger fraction ($>40$\%;
  as  it might  be  expected) for  low-\mh\  centrals}.  Despite  this
heterogeneity, as shown in Figs.~\ref{fig:CEN_SL} and~\ref{fig:H_CEN},
low-  and   high-mass  centrals  show   remarkably  different  stellar
population   properties,  reinforcing   the   conclusion  that   these
differences are  driven by the ``global'' environment  (i.e.  the halo
mass) where galaxies reside.

\begin{figure}
\begin{center}
\leavevmode
\includegraphics[width=8cm]{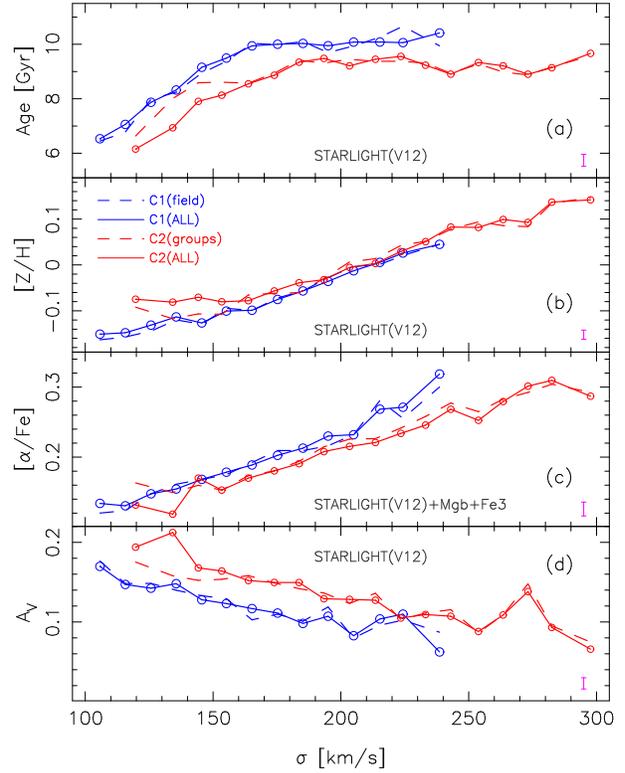}
\end{center}
\caption{ Trends of stellar population properties for stacked
  spectra of low-\mh\ centrals (C1) classified as ``field'' galaxies,
  and high-\mh\ centrals (C2) classified as ``group'' galaxies
  according to Paper III \citep[based on the group catalogue
    of][]{Berlind:06}. Solid curves show the trends for the original
  samples of centrals, C1 and C2, used in this paper, i.e. as in 
  Fig.~\ref{fig:CEN_SL}.  }

%\contcaption{.}
\label{fig:CEN_SL_BERLIND}
\end{figure}

{ To further analyze this aspect, we have constructed stacked
  spectra for two subsamples of central ETGs, consisting of (i) C1
  centrals classified as field galaxies according to Paper III
  ($N_{ETGs}=4,380$), and (ii) C2 centrals classified as group
  galaxies from Paper III ($N_{ETGs}=2,349$).
  Fig.~\ref{fig:CEN_SL_BERLIND} compares the trends of stellar
  population properties with $\sigma$ for these subsamples with those
  for C1 and C2.  While for C1 there is no significant change when
  selecting only ``field'' galaxies, selecting only C2 ETGs classified
  as ``group'' from Paper III produces some changes, in that (i) the
  difference in metallicity between C1 and C2 tends to disappear (see
  blue and red dashed curves in panel b of
  Fig.~\ref{fig:CEN_SL_BERLIND}), and (ii) at low $\sigma$
  ($\simlt 150$\,\kms ) the difference in age, \afe , and
  \av\ becomes smaller than that between the whole samples of C1 and
  C2 ETGs.  This test indicates that the dependence of stellar
  population properties on \mh\ holds up against a different definition
  of environment. The amplitude of the differences in stellar population 
  properties between isolated and group ETGs is more significant when we 
  add, to the C2 ``group'' galaxies of Paper III, those C2 centrals labelled 
  ``field'' or ``unclassified'' in the same paper, likely residing in 
  low-multiplicity groups, that the virial analysis of Paper III has not 
  been able to measure.} 

Since our samples  of satellites mostly consist of  group galaxies, we
can     compare     our     findings,    in     Figs.~\ref{fig:SAT_SL}
and~\ref{fig:H_SAT},  with those of  previous studies  contrasting the
properties of ETGs as a function of local galaxy density.  Indeed, the
only environmental  effect we  detect for samples  S1, S2, and  S3, is
that at low-$\sigma$, ETGs  at large group-centric projected distances
(sample S3)  tend to have younger stellar  populations than satellites
in  the central regions.   This is  consistent with~\citet{Rogers:10},
and  the general finding  that low-mass  ETGs residing  in low-density
environments  have  younger stellar  populations  than  those at  high
density.  { On the  other hand,~\citet{Bernardi:09} found that BCGs
  in  groups are  $\sim$0.5--1\,Gyr older  than their  satellites (see
  also~\citealt{Pasquali:10}).  We  notice  that  this result  is  not
  inconsistent  with our  findings as  (i)  we are  not comparing  the
  properties of satellites with their  own centrals (samples C1 and C2
  have different \mh\  than samples S1, S2, and  S3), but, rather, the
  two populations individually  as a whole; and (ii)  we are analyzing
  only ETGs rather than galaxies of all morphological types.}

Regarding metallicity, we also find evidence that low-$\sigma$ ETGs in
sample S3 (i.e.  galaxies in ``low-density'' regions) have slightly
higher \zh\ than satellites in the cluster cores (samples S1 and S2).
This is consistently seen in the \hbo--\mgfep\ diagram
(Fig.~\ref{fig:H_SAT}), where the trend for sample S3 (green curve) is
shifted rightwards (i.e.  towards higher \zh ) with respect to the
trends for samples S1 and S2.  This finding is consistent with
previous studies~\citep{Thomas:05, deLaRosa:07, Clemens:09, Zhu:10},
as well as our inference from the environmental dependence of the
Fundamental Plane relation of ETGs (Paper III). One should notice that
some studies did not detect any environmental dependence of
metallicity (e.g.~\citealt{Berlind:06, Annibali:07, Harrison:11}),
while others~\citep{Gallazzi:06} found ETGs in low-density
environments to be less metal-rich than those at high density.  These
discrepancies likely arise both because of different definitions of
environment, and because of the intrinsically small differences found
in the stellar population properties among different environments.

Our  new results  can  also be  compared  with those  of PGF10,  where
low-mass satellites, in low-mass  halos, were found to feature younger
ages than those  at increasing \mh .  PGF10  interpreted the age trend
as a  result of  satellites in today  more, relative to  less, massive
halos   having  been  accreted   at  earlier   epochs.  As   shown  in
Fig.~\ref{fig:SAT_SL}, comparing the low-$\sigma$ behaviour of samples
S1 and S2, our results are consistent with this finding.

\begin{table}
\small
\begin{minipage}{80mm}
 \caption{Percentage  of galaxies  in  each sample  of ETGs  (col.~1),
   classified  as field  and group  systems, as  well  as unclassified
   objects, according to Paper III (cols.~2, ~3, and~4, respectively).
 } \centering
  \begin{tabular}{c|c|c|c}
 \hline
sample & field & group & un-classified\\
(1) & (2) & (3) & (4) \\
\hline
C1 &   42\%  &  27\% & 31\% \\
C2 &   32\%  &  47\% & 21\% \\
S1 &   12\%  &  62\% & 26\% \\
S2 &    0\%  &  84\% & 16\% \\
S3 &    3\%  &  78\% & 19\% \\
\hline
\end{tabular}
\label{tab:field_group}
\end{minipage}
\end{table}

%%%%%%%%%%%%%%%%%%%%%%%%%%%%%%%%%%%%%%%%%%%%%%%%%%%%%%%%%%%%%%%%%
\section{Summary and Discussion}
\label{sec:discussion}

In the  present work, we have  studied how the  stellar populations of
ETGs depend  on the ``global'' environment, characterised  by the mass
of the host halo.  Since a key driver of the formation history of ETGs
is  the central  velocity dispersion  ($\sigma$, a  ``local'' quantity
mainly dependent  on the evolution  of the gravitational  potential of
the galaxy), we perform the comparisons regarding environment at fixed
$\sigma$. We have performed a  stellar population study of central and
satellite ETGs, using  72 stacked SDSS spectra, covering  a wide range
in velocity dispersion: $100\!<\!\sigma\!<\!310$\,\kms . 
{ The satellite
sample  is split  into  three subsamples,  namely  galaxies residing  in
the central regions ($<0.5 \, R_{200}$) of ``groups''  (with halo  mass \lmh$<\!14$)  and  ``clusters'' (\lmh$\ge
14$), respectively, and those residing in the outskirts ($\ge 0.5 \, R_{200}$) of all groups/clusters 
(regardless of \mh )}. Central ETGs  are binned into ``isolated'' systems
(\lmh$<\!12.5$)    and   those    residing   in    galaxy   ``groups''
(\lmh$\ge\!12.5$).  Our results are summarised as follows:

\begin{description}
 \item[{\it Global trends:}] The  stellar age, metallicity (\zh ), and
   \afe\   increase   with    velocity   dispersion,   regardless   of
   environment. The slopes of the \zh\ and \afe\ -- $\sigma$ relations
   are  consistent  with  previous  studies  (e.g.~\citealt{Trager:00,
     Bernardi:03,    Thomas:05,     Smith:07,    Zhu:10,    Thomas:10,
     Harrison:11}). The  age increases  with $\sigma$ in  low velocity
   dispersion systems ($\sigma< 150$--$200$\,\kms ), whereas at higher
   velocity dispersion,  it either remains constant  or increases with
   $\sigma$,  depending  on the  method  used  to  derive the  stellar
   population  properties. ETGs at  high $\sigma$  tend to  have lower
   internal extinction than those at  low $\sigma$. This trend is more
   significant in centrals than satellites.
 \item[{\it Centrals:}]  The stellar population  properties of central
   ETGs depend significantly on global environment.  At fixed velocity
   dispersion, the centrals of galaxy groups have younger ages (at all
   values  of  $\sigma$),   higher  metallicites  (especially  at  low
   $\sigma$), and  lower \afe\ than ``isolated''  centrals (i.e. those
   in lower mass halos).
 \item[{\it Satellites:}] In contrast  to the centrals, no significant
   environmental dependence of  stellar population properties is found
   in  satellite  ETGs, except  for  systems  at  the lowest  velocity
   dispersions  ($\sigma\!<\!130$\,\kms),  where  satellites  residing
   either in  low-mass halos or in  the outskirts of  the more massive
   groups, seem to  have younger ages than satellite  ETGs residing in
   massive halos. These results are consistent with those of PGF10. In
   particular, the fact that,  at low $\sigma$, satellites in low-mass
   halos have younger ages than  those with high \mh\ can be explained
   by  a later infall  into the  host halo,  followed by  quenching of
   their      star-formation      by      environmental      processes
   (e.g. strangulation).
 \end{description}

\begin{figure}
\begin{center}
\leavevmode
\includegraphics[width=8.cm]{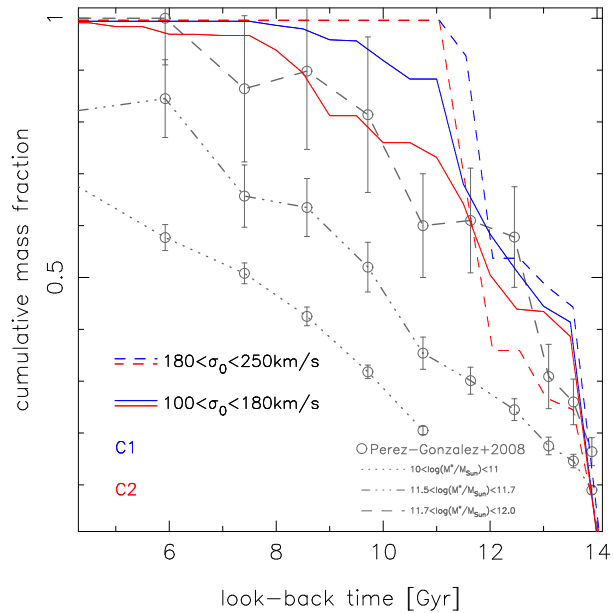}
\end{center}
\caption{The stellar  mass formation  histories of the  populations in
  central ETGs is  shown as a function of  look-back time. Notice that
  the  plot is  limited  to epochs  older  than $\sim  4$\,Gyr as  the
  stellar mass fractions are identically equal to one below this value
  (i.e.  negligible late star formation in our sample).  The solid and
  dashed   curves   are   the   median   trends  for   ETGs   in   low
  ($\sigma<180$\,\kms)   and    high   ($\sigma>180$\,\kms)   velocity
  dispersion bins,  respectively.  Blue and red  colours correspond to
  the  centrals in low-  and high-mass  halos, respectively,  with the
  same colour coding as in Figs.~\ref{fig:CEN_SL} and~\ref{fig:H_CEN}.
  The figure shows that stellar  populations in central ETGs in groups
  (high-mass  halos)  are  formed  at  a  lower  pace  than  those  in
  ``isolation''  (i.e. ETGs  residing in  low-mass  halos), consistent
  with    the   results   of    our   stellar    population   analysis
  (Fig.~\ref{fig:CEN_SL}).  Also,  notice that star  formation is more
  extended  towards  lower velocity  dispersion.   For comparison,  we
  overplot the results  from~\citet[][grey]{PG:2008} for three stellar
  mass  bins, covering  the entire  stellar mass  range of  our sample
  (grey labels). }
%\contcaption{.}
\label{fig:mass_frac_cen}
\end{figure}

 The new main result of the  present work is the dependence of stellar
 population  properties  of central  ETGs  on  environment. {\sl  What
   causes younger  ages in  the central ETGs  of more  massive halos?}
 Since we analyse  stacked spectra, these young ages  may arise either
 because of a more  prolonged star-formation in the average population
 of  high-\mh\ galaxies,  or because  of  a larger  fraction of  these
 galaxies   with    young   stellar   populations.     For   instance,
 \citet{Thomas:10} found that while the  bulk of ETGs have $Age$, \zh,
 and  \afe\  vs.  $\sigma$  trends  independent  of local  environment
 (i.e. galaxy density), low-density  regions feature a larger fraction
 of ETGs with  stellar populations ``re-juvenated'' by a  few Gyr.  In
 Fig.~\ref{fig:mass_frac_cen}, we present  the cumulative stellar mass
 formation  histories (SMFHs)  for low-  and high-\mh\  centrals.  The
 SMFHs  are obtained  from the  {\tt  STARLIGHT} fits  to the  stacked
 spectra  (see  Sec.~\ref{sec:age_zh}), as  detailed  in our  previous
 works~\citep{delaRosa:11,  Trevisan:12}.  In  the Figure,  we average
 the results for stacks  at low (i.e. $100\!<\!\sigma\!<\!180$\,\kms )
 and    high   (i.e.    $180\!<\!\sigma\!<\!250$\,\kms    )   velocity
 dispersion. Centrals show smoothly  declining SMFHs, with the bulk of
 the  stellar mass  formed at  a lookback  time $\simgt  6$\,Gyr.  For
 centrals  in more  massive groups  (red  curves in  the Figure),  the
 stellar  component formed  over a  more extended  time scale  than in
 ``isolated'' (low  halo mass) centrals  (blue curves). This  is fully
 consistent  with the  results shown  in  Fig.~\ref{fig:CEN_SL}, which
 suggested that high-\mh\ centrals  have younger ages, and lower \afe,
 than  low-\mh\  centrals.   Since   Fe  mainly  originates  from  the
 explosion  of type  Ia supernov\ae,  over  a larger  time scale  than
 $\alpha$ elements  (mostly produced in short-lived  massive stars), a
 lower  \afe\ does  in  fact  imply a  more  prolonged star  formation
 history,      consistent      with      Fig.~\ref{fig:mass_frac_cen}.
 Quantitatively, we find  that at high $\sigma$, centrals  in low mass
 halos  have \afe\ $\sim  0.025$\,dex higher  (on average)  than their
 counterparts  in more  massive halos  (Fig.~\ref{fig:CEN_SL}).  Using
 eq.~2  of~\citet{delaRosa:11}, such  a difference  translates  into a
 half-mass formation time difference of $\sim 0.4$\,Gyr, qualitatively
 consistent  with the  different shapes  of  the red  and blue  dashed
 curves  in  Fig.~\ref{fig:mass_frac_cen}.    For  centrals  with  low
 $\sigma$, we  also find a more  prolonged SMFH in  more massive halos
 (compare the  red and blue solid  curves), despite the  fact that the
 difference in  \afe\ between  low- and high-  \mh\ samples is  not as
 significant    as   at    high    $\sigma$   (Fig.~\ref{fig:CEN_SL}).
 Fig.~\ref{fig:mass_frac_cen}  also  compares  our  SMFHs  with  those
 obtained  by~\citet{PG:2008},  from  a  sample  of  $28,000$  objects
 extended to all morphological types, at $0<z<4$.  Because of the high
 star-formation      efficiency~\citep[see,      e.g.,][]{Ferreras:09,
   Trevisan:12},  ETGs have  shorter  formation time  scales than  the
 overall population of galaxies,  with the difference becoming smaller
 towards massive galaxies,  as expected by the fact  that most massive
 galaxies are early-type systems.

Within the current paradigm  of galaxy formation, central galaxies can
accrete gas from the hot-gas  reservoir of their host halo \citep[see,
  e.g.,][]{Croton:06}.  In  the most  massive halos, this  hot-mode of
star-formation  is believed  to  be suppressed  by feedback  processes
(e.g.  AGN activity), in  order to  explain the  old stellar  ages and
low-level of recent star-formation in massive galaxies at low redshift
\citep{delucia:06}. The more prolonged  SMFHs of centrals in high-mass
halos  might be  due  to a  stronger  accretion of  hot  gas from  the
surrounding halo. However,  the gas should also cool  down to sit into
the  central galaxy  itself to  form  a disc  (preserving its  angular
momentum). Since our sample  is selected to minimise the contamination
from systems with a disk component (see Sec.~\ref{sec:data}), we would
expect  this scenario  to be  ruled  out.  Instead,  we interpret  our
findings as the result of gas-rich galaxy-galaxy interactions, between
centrals  residing  in  present-day  groups,  and  the  population  of
infalling satellites. These  interactions are certainly more important
in  groups, than  in ``isolated''  ellipticals (where,  by definition,
there  are fewer satellites,  see Fig.~\ref{fig:hmass_hist}).  In this
framework,  dissipative  interactions should  have  been an  essential
ingredient of the mass accretion history of central ETGs in groups. In
fact, major (minor) ``dry'' (dissipation-less) mergers are expected to
increase      (not      change)      the      velocity      dispersion
\citep{Hilz:12}. Assuming that the trends for ``isolated'' centrals at
low redshift are representative of those of the progenitors of central
ETGs in more  massive groups today, major dry mergers  would lead to a
horizontal shift  (see Fig.~\ref{fig:CEN_SL}) towards  higher $\sigma$
of {\sl  all} stellar population trends regarding  central ETGs. While
the  trends  for $Age$,  \afe  , and  \av\  are  consistent with  this
scenario,  the trends  for metallicity  are  not. On  the other  hand,
simulations   suggest  that  gas-rich   interactions  can   produce  a
significant dilution  of metallicity only  during the first  phases of
the  interaction,  especially for  intense  levels of  star-formation,
after  which   enrichment  becomes   the  dominant  effect,   so  that
metallicity is expected  to be larger in the final  system than in the
progenitors~\citep{Montuori:10}. This mechanism  goes in the direction
of increasing  the metallicity of  centrals in more massive  halos, as
observed  in Fig.~\ref{fig:CEN_SL}.   Therefore, the  results  of this
work point  consistently to a  picture whereby central ETGs  in groups
evolve through a higher number of gas-rich galaxy-galaxy interactions,
resulting  in slightly  more  extended star  formation histories  with
respect  to  ``isolated''  ETGs  -- although  the  ``local''  physics,
parameterised as central velocity  dispersion, remains the main driver
of  star formation  in  ETGs, a  result  that is  consistent with  the
properties   of   star    forming   galaxies   as   well   \citep[see,
  e.g.,][]{Wijesinghe:12}.

It  is   also  worth  emphasising  that  the   subtle  but  consistent
differences  found  between the  stellar  populations  in central  and
satellite  ETGs confirm  that  a central/satellite  split  is the  key
property  that  allows  us  to  understand in  detail  the  effect  of
environment on galaxy formation.

%%%%%%%%%%%%%%%%%%%%%%%%%%%%%%%%%%%%%%%%%%%%%%%%%%%%%%%%%%%%%%%%%
\section*{Acknowledgements}
{ We would like to thank the anonymous referee who helped us to 
significantly improve this manuscript.}
AG  acknowledges support  from  the European  Union Seventh  Framework
Programme (FP7/2007-2013)  under grant  agreement n. 267251.   We have
used    data   from    the    Sloan   Digital    Sky   Survey    ({\tt
  http://www.sdss.org/collaboration/credits.html}). We  have used data
from the 4th data release of the UKIDSS survey \citep{Law07}, which is
described  in  detail in  \citet{War07}.   Funding  for  the SDSS  and
SDSS-II  has been  provided by  the Alfred  P.  Sloan  Foundation, the
Participating Institutions, the  National Science Foundation, the U.S.
Department   of   Energy,   the   National   Aeronautics   and   Space
Administration, the  Japanese Monbukagakusho, the  Max Planck Society,
and the Higher Education Funding Council for England.

%%%%%%%%%%%%%%%%%%%%%%%%%%%%%%%%%%%%%%%%%%%%%%%%%%%%%%%%%%%%%%%%%%%%%
\appendix

%%%%%%%%%%%%%%%%%%%%%%%%%%%%%%%%%%%%%%%%%%%%%%%%%%%%%%%%%%%%%%%%%
\section{Running {\tt STARLIGHT} with $\alpha$-enhanced SSPs}
\label{app:afe_ssps}

\begin{figure}
\begin{center}
\leavevmode
\includegraphics[width=8cm]{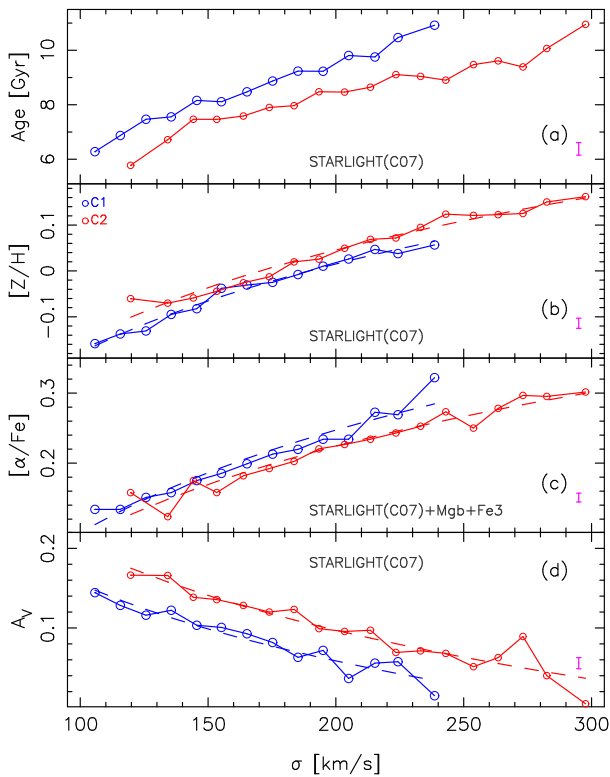}
\end{center}
\caption{Same   as   Fig.~\ref{fig:CEN_SL}   but  deriving   age   and
  metallicity       with       $\alpha$-enhanced      SSP       models
  from~\citet{Cervantes07}.  Notice that \afe\  is estimated  from the
  solar-scale proxy,  \afep, as in  Fig.~\ref{fig:CEN_SL}, whereas age
  and  metallicity  are  obtained  by  running  {\tt  STARLIGHT}  with
  $\alpha$-enhanced SSPs.  }
%\contcaption{.}
\label{fig:CEN_SL_CE}
\end{figure}

\begin{figure}
\begin{center}
\leavevmode
\includegraphics[width=8cm]{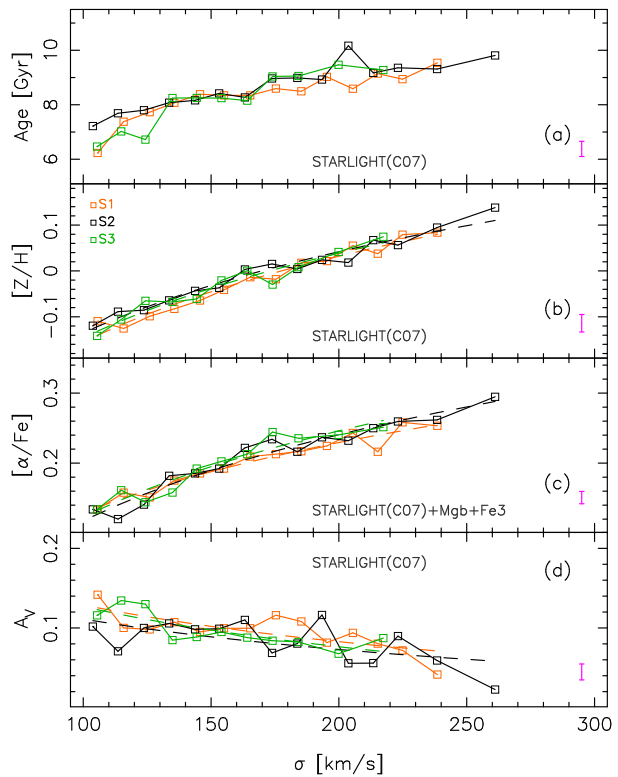}
\end{center}
\caption{Same  as Fig.~\ref{fig:CEN_SL_CE}  but  for satellite  rather
  than central ETGs.  }
%\contcaption{.}
\label{fig:SAT_SL_CE}
\end{figure}

We test the effect of using solar-scaled models on the derived stellar
population  properties  of  ETGs  by  comparing  our  results  with  a
preliminary set of  $\alpha$-enhanced MILES (hereafter $\alpha$-MILES)
SSP  models constructed  by Cervantes  et. al.~(2007),  based  on both
empirical  and  theoretical stellar  libraries.   The empirical  MILES
library (S\'anchez-Blazquez et al.  2006) contains spectra of stars in
the solar neighbourhood, but  mostly lacks bulge-stars, with non-solar
abundance ratios.  The  $\alpha$-MILES models complement the empirical
library, in the non-solar abundance regime, with the synthetic library
of  Coelho  et  al.~(2007),  consisting of  high-resolution  synthetic
stellar  spectra   covering  a  wide  range   of  stellar  atmospheric
parameters.  The  Coelho et al.~(2007)  library covers both  solar and
alpha-enhanced   mixtures   over  a   wide   wavelength  range,   from
$3000$\AA\ to 1.4\,$\mu$m, superseding  previous versions of Barbuy et
al.     (2003),    in    the    wavelength    range    $4600-5600$\AA,
and~\citet{Zwitter:2004}, in the  range $7653-8747$\AA.  The resulting
$\alpha$-enhanced models  consist of  SEDs covering the  same spectral
range ($3525-7500$ \AA), with the same spectral resolution ($2.3$\AA),
as  the (nearly  solar-scaled) MILES  models used  in this  paper.  As
shown in  Paper IV, the preliminary $\alpha$-enhanced  models are more
effective  at  describing  the   spectra  of  massive  ETGs  than  the
solar-scaled  models,  in   particular  concerning  spectral  features
strongly sensitive  to \afe  , such as  Mg lines.   The $\alpha$-MILES
models consist of $1,170$ SSPs, corresponding to: twenty-six ages from
$1$ to  $18$\,Gyr; five metallicities from \zh$=-1.28$  to $+0.2$, and
nine value of  \afe, from $-0.2$ to $+0.6$.  With  this set of models,
we have run  {\tt STARLIGHT} on our stacked spectra,  using a basis of
$224$ $\alpha$-MILES  SSPs~(see Sec.~\ref{sec:age_zh}), with  14 steps
in age,  from $1$  to $14$\,Gyr, four  steps in \zh\  ($-0.7$, $-0.4$,
$0$, and $+0.2$), and four  steps in \afe\ ($0$, $0.2$, $0.4$, $0.6$).
For  each stack, we  compute age  and metallicity  from $\alpha$-MILES
best-fitting    ({\tt   STARLIGHT})    results,   as    described   in
Sec.~\ref{sec:age_zh}  (i.e. from  Eq.~\ref{eq:lum_weight_SP}). Notice
that one could use the  $\alpha$-MILES {\tt STARLIGHT} results also to
compute    luminosity-weighted    estimates    of    \afe\    (through
Eq.~\ref{eq:lum_weight_SP}).   However,  we  prefer  to  rely  on  our
solar-scale   proxy  \afep\   (Sec.~\ref{sec:alpha}),   obtained  with
solar-scaled   single-SSP  MILES   models,   as  such   SSP-equivalent
\afe\  estimates can be  related more  directly to  the star-formation
time scale of ETGs~\citep{delaRosa:11}.  Thus, we re-compute \afe\ for
each   stacked    spectrum   using    the   same   approach    as   in
Sec.~\ref{sec:alpha},  but  using  luminosity-weighted ages  from  the
$\alpha$-MILES   {\tt   STARLIGHT}  fits.    Figs.~\ref{fig:CEN_SL_CE}
and~\ref{fig:SAT_SL_CE}   are  the   same   as  Figs.~\ref{fig:CEN_SL}
and~\ref{fig:SAT_SL},   respectively,  but   for   stellar  population
parameters  from $\alpha$-MILES (instead  of solar-scaled  MILES) {\tt
  STARLIGHT}  fits.    In  general,  using   $\alpha$-MILES  leads  to
consistent  results with  respect to  the MILES  models,  although the
trends   of  $Age$   and   \av\  with   $\sigma$   are  steeper   (see
Secs.~\ref{sec:results}    and~\ref{sec:comps},   respectively).    In
particular,  the main  results of  the present  work --  regarding the
environmental  comparison of  satellite  and central  ETGs  -- do  not
depend on the adopted stellar population models.

\section{Comparison with independent parameter estimates on individual galaxies}
\label{app:compAG}

\begin{figure}
\begin{center}
\leavevmode
\includegraphics[width=8cm]{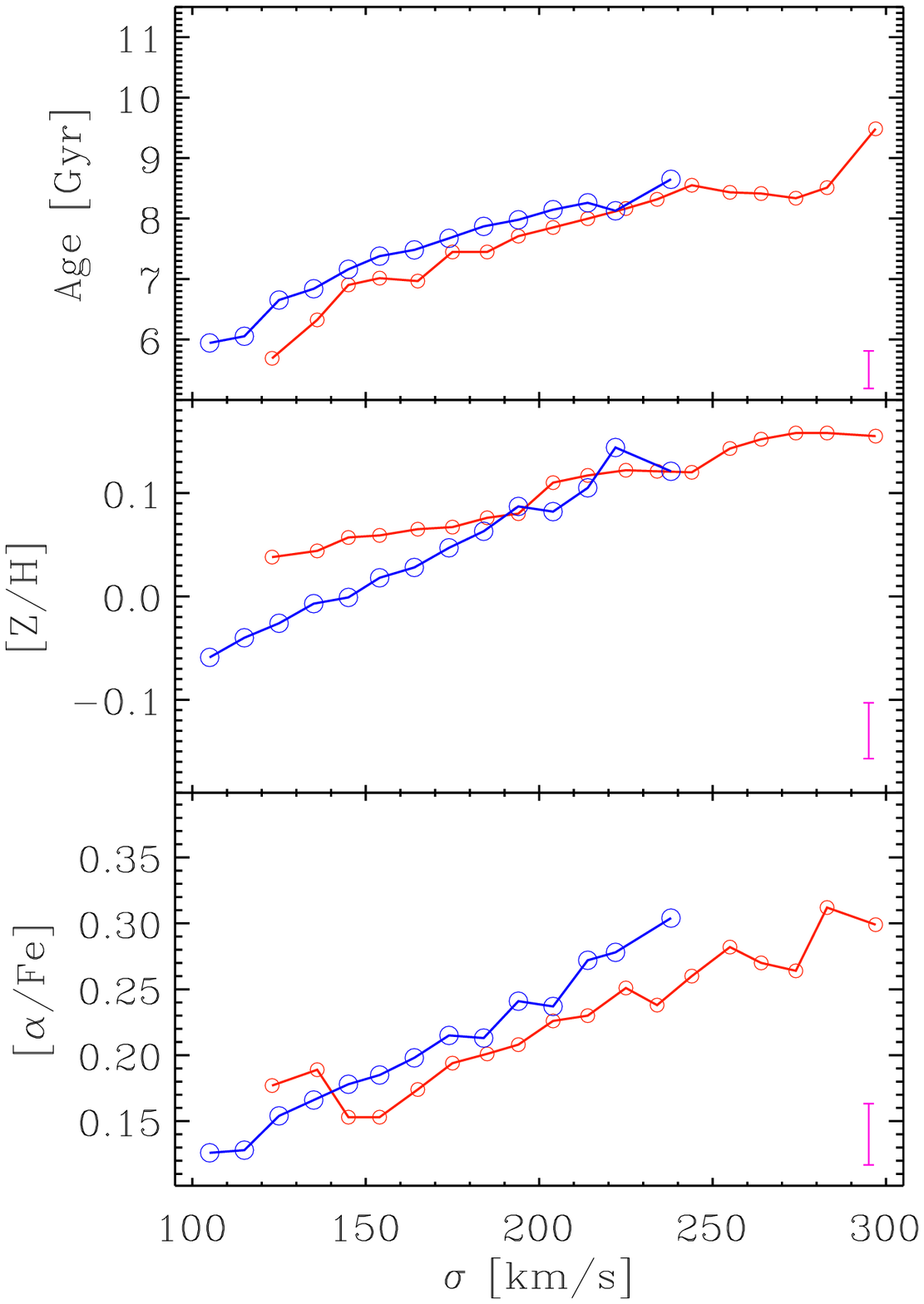}
\end{center}
\caption{Luminosity-weighted    age,     stellar    metallicity    and
  [$\alpha$/Fe]  as  a function  of  velocity  dispersion for  central
  galaxies with \lmh$> 12.5$ (red)  and with \lmh$< 12.5$ (blue). Ages
  and metallicities are derived as in \citet{Gallazzi:06} by comparing
  a  set of absorption  features with  a Monte  Carlo library  of star
  formation histories based  on the solar-scaled \citet{BC:03} models.
  The   [$\alpha$/Fe]  estimates  are   derived  from   the  empirical
  definition  of \citet{Gallazzi:06},  $\Delta({\rm  Mgb}/\langle {\rm
    Fe}\rangle)$, calibrated  with the \citet{TMB:03}  models.  In all
  panels, magenta error bars  in the lower-right corners correspond to
  the maximum  values of $1$~sigma  uncertainties on median  values of
  stellar population properties among different data-points.  }
%\contcaption{.}
\label{fig:CEN_AG}
\end{figure}

\begin{figure}
\begin{center}
\leavevmode
\includegraphics[width=8cm]{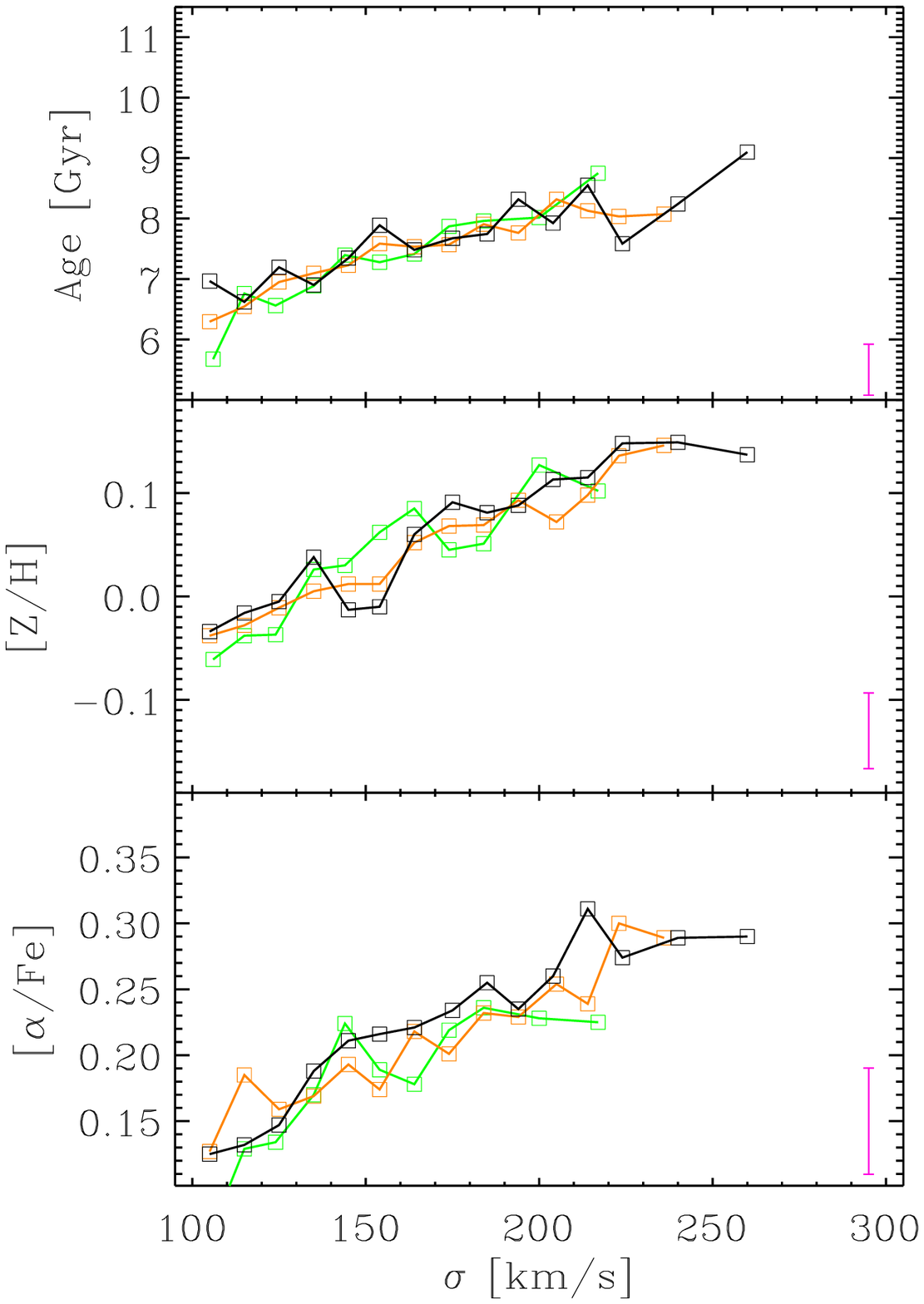}
\end{center}
\caption{  Same as Fig.~\ref{fig:CEN_AG},  but for  satellite galaxies
  (same colour-coding as in Fig.~\ref{fig:CEN_SL}).  }
%\contcaption{.}
\label{fig:SAT_AG}
\end{figure}

To  further validate  the  robustness of  our  results, in  particular
against differences in stellar  population modelling, we have compared
them   with  those   obtained  from   the   analysis  of   a  set   of
\afe-independent  absorption features of  SDSS early-type  galaxies by
\cite{Gallazzi:06}.   In  summary,  the luminosity-weighted  ages  and
stellar metallicities are derived via a Bayesian approach by comparing
the    strengths    of    $\rm    D4000_n$,   $\rm    H\beta$,    $\rm
H\delta_A+H\gamma_A$, $\rm [Mg_2Fe]$, $\rm [MgFe]^\prime$ with a Monte
Carlo library  of star formation  histories based on  the \cite{BC:03}
population synthesis models. Estimates  of element abundance ratio are
based  on  the  empirical  estimator  defined  in  \cite{Gallazzi:06},
$\Delta$(Mgb/$\langle$Fe$\rangle$),  which   is  the  excess   of  the
observed   Mgb/$\langle$Fe$\rangle$  feature   with  respect   to  the
solar-scaled  model that  best fits  the above  defined  indices. This
estimator is  calibrated into  [$\alpha$/Fe] using the  predictions of
\cite{TMB:03}  models  (Gallazzi et  al.,  in  prep.).   Note that  in
contrast  to  the rest  of  this  paper,  these parameters  have  been
estimated   for   {\sl    individual}   SDSS   galaxies.    For   each
central/satellite  galaxy subsample and  for each  velocity dispersion
bin, we  take the median luminosity-weighted  age, stellar metallicity
and \afe\  of all  the galaxies that  contribute to  the corresponding
stacked   spectrum.   In   analogy  with   Figs.~\ref{fig:CEN_SL}  and
\ref{fig:SAT_SL},      we      show     in      Figs.~\ref{fig:CEN_AG}
and~\ref{fig:SAT_AG}  the results  obtained for  central  galaxies and
satellite galaxies,  respectively.  We  notice a few  differences with
respect to Figs.~\ref{fig:CEN_SL}  and \ref{fig:SAT_SL}. In particular
stellar age does not show a flattening at high velocity dispersion but
it    keeps    on     increasing    with    $\sigma$,    similar    to
Fig.~~\ref{fig:CEN_SL_CE}   (derived  from   a   preliminary  set   of
$\alpha$-enhanced models).   The slope is  however shallower, reaching
ages about 1\,Gyr younger at high velocity dispersion than those shown
in  the previous  figures.  Despite  these differences,  the agreement
with   the    results   discussed   in    Figs.~\ref{fig:CEN_SL}   and
\ref{fig:SAT_SL} is  remarkable.  In particular, we  also confirm with
these independent estimates, that  i) central galaxies in larger halos
are younger,  more metal-rich and less  $\alpha$-enhanced than central
galaxies  in lower-mass  halos,  and ii)  the  stellar populations  of
satellite  early-type galaxies do  no show  any systematic  trend with
environment. This comparison also reassures us about the similarity of
parameters obtained  from our stacked  spectra with respect  to median
parameter values of galaxies contributing  to each bin as derived from
individual spectra.

%%%%%%%%%%%%%%%%%%%%%%%%%%%%%%%%%%%%%%%%%%%%%%%%%%%%%%%%%%%%%%%%%
\section{Correlated uncertainties of stellar population parameters.}
\label{app:afe_deg}

\begin{figure*}
\begin{center}
\leavevmode
\includegraphics[width=14cm]{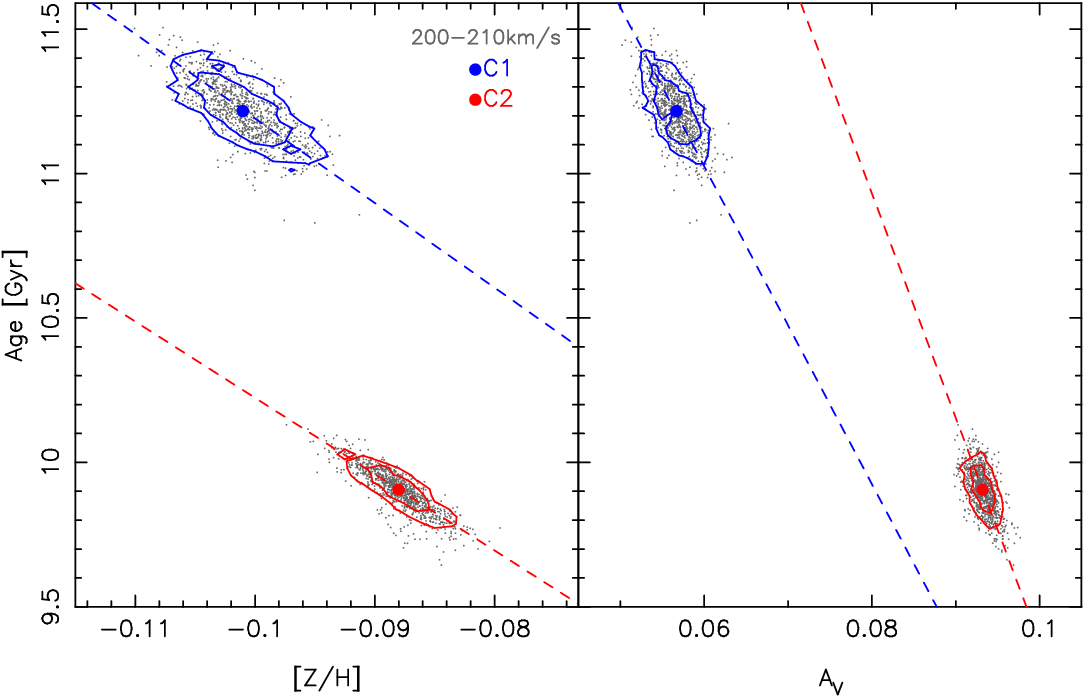}
\end{center}
\caption{The  effect of  correlated uncertainties  on $Age$,  \zh, and
  \av\ is shown for two representative stacked spectra from samples C1
  and  C2,  with  velocity  dispersion  $\sigma\sim  200$--$210$\,\kms
  .  Grey points  are obtained  by shifting  the flux  values  of each
  spectrum according  to the observed  uncertainties, and re-measuring
  $Age$, \zh,  and \av\ with  {\tt STARLIGHT}, accordingly.   From the
  re-measured   values  (grey  dots),   we  estimate   1-$\sigma$  and
  2-$\sigma$  confidence contours  in the  $Age$ vs.  \zh\  (left) and
  $Age$ vs. \av\ (right) diagrams, for the C1 and C2 spectra (blue and
  red  colours,  respectively).  Dashed  lines show  the  direction  of
  correlated  uncertainties.  Notice  that the  degeneracy  directions
  implied for C1 and C2 are almost parallel.}
%\contcaption{.}
\label{fig:corr_sps}
\end{figure*}

We  illustrate  the  effect  of correlated  uncertainties  on  stellar
population parameters by considering  two stacked spectra from samples
C1 and  C2, respectively, both at  a velocity dispersion  in the range
$\sigma\sim 200$--$210$\,\kms . For each spectrum, we performed $1000$
iterations,  where $Age$,  \zh,  and \av\  are  re-measured with  {\tt
  STARLIGHT} on  spectra where the  flux values are  randomly modified
according  to  their  uncertainties  (see  Sec.~\ref{sec:sp_pars}  for
details).  Fig.~\ref{fig:corr_sps}  shows   the  distribution  of  the
measurements  for  both spectra,  in  the  $Age$  vs. \zh\  and  $Age$
vs.  \av\  diagrams,  including   the  1  and  2\,$\sigma$  confidence
levels.  As expected, measurement  errors tend  to shift  the fiducial
values in the plots  (filled circles) along anti-correlated directions
of   $Age$--\zh\   and   $Age$--\av\   (see  dashed   lines   in   the
Figure). However, the directions  of correlated errors, between C1 and
C2 stacks, are almost parallel, i.e. the difference of $Age$, \zh, and
\av\  between samples C1  and C2  do not  arise because  of correlated
uncertainties. Would this  have been the case, we  would have measured
much  larger  differences in  metallicity  with  respect  to those  we
measure in  the $Age$  parameter, as this  would be required  to shift
horizontally the red dot  in the left panel of Fig.~\ref{fig:corr_sps}
towards the  blue dashed curve.  This result gives further  support to
the fact  that the environmental  differences detected in  the present
work, and in particular those  for central ETGs, arise neither because
of  a  degeneracy  of   stellar  population  properties,  nor  from  a
particular marginalisation of the uncertainties in the parameter space
explored.

%%%%%%%%%%%%%%%%%%%%%%%%%%%%%%%%%%%%%%%%%%%%%%%%%%%%%%%%%%%%%%%%%

\label{lastpage}

\end{document}